\PassOptionsToPackage{table,xcdraw}{xcolor}
\documentclass[sigconf]{acmart}
\usepackage{subcaption}
\usepackage{csquotes}
\usepackage[capitalise,noabbrev,nameinlink]{cleveref}  
\usepackage{colortbl}
\usepackage{booktabs}  
\usepackage{tabularx}  
\usepackage{makecell}  

\definecolor{tablebg}{rgb}{0.92,0.92,0.92}
\definecolor{gamesexp-blue}{HTML}{005AB5}
\definecolor{gamesexp-red}{HTML}{DC3220}

\newcommand{\nowordbreak}[1]{\mbox{#1}} 

\definecolor{revisions}{HTML}{000000}  

\newcommand{\rev}[1]{\textcolor{revisions}{#1}}  

\AtBeginDocument{%
  \providecommand\BibTeX{{%
    \normalfont B\kern-0.5em{\scshape i\kern-0.25em b}\kern-0.8em\TeX}}}

\setcopyright{acmlicensed}
\copyrightyear{2024}
\acmYear{2024}
\acmDOI{10.1145/3613904.3642077}
\acmISBN{979-8-4007-0330-0/24/05}

\acmConference[CHI '24]{Proceedings of the CHI Conference on Human Factors in Computing Systems}{May 11--16, 2024}{Honolulu, HI, USA}
\acmBooktitle{Proceedings of the CHI Conference on Human Factors in Computing Systems (CHI '24), May 11--16, 2024, Honolulu, HI, USA}

\begin{document}

\title[Not All the Same: Similarity Estimation in Tile-Based Video Games]{Not All the Same: Understanding and Informing Similarity Estimation in Tile-Based Video Games}

\author{Sebastian Berns}
\orcid{0000-0001-9555-5954}
\affiliation{%
  \institution{Queen Mary University of London}
  \country{United Kingdom}
}
\email{s.berns@qmul.ac.uk}

\author{Vanessa Volz}
\orcid{0000-0002-4919-4374}
\affiliation{%
  \institution{modl.ai}
  \country{Denmark}
}

\author{Laurissa Tokarchuk}
\orcid{0000-0002-3118-5031}
\affiliation{%
  \institution{Queen Mary University of London}
  \country{United Kingdom}
}

\author{Sam Snodgrass}
\orcid{0009-0007-6554-0808}
\affiliation{%
  \institution{modl.ai}
  \country{Denmark}
}

\author{Christian Guckelsberger}
\orcid{0000-0003-1977-1887}
\affiliation{%
  \institution{Aalto University}
  \country{Finland}
}
\affiliation{%
  \institution{Queen Mary University of London}
  \country{United Kingdom}
}

\renewcommand{\shortauthors}{Berns, et al.}

\begin{abstract}
Similarity estimation is essential for many game AI applications, from the procedural generation of distinct assets to automated exploration with game-playing agents. While similarity metrics often substitute human evaluation, their alignment with our judgement is unclear. Consequently, the result of their application can fail human expectations, leading to e.g. unappreciated content or unbelievable agent behaviour. We alleviate this gap through a multi-factorial study of two tile-based games in two representations, where participants (N=456) judged the similarity of level triplets. Based on this data, we construct domain-specific perceptual spaces, encoding similarity-relevant attributes. We compare 12 metrics to these spaces and evaluate their approximation quality through several quantitative lenses. Moreover, we conduct a qualitative labelling study to identify the features underlying the human similarity judgement in this popular genre. Our findings inform the selection of existing metrics and highlight requirements for the design of new similarity metrics benefiting game development and research. 
\end{abstract}



\begin{teaserfigure}
  \includegraphics[width=\textwidth]{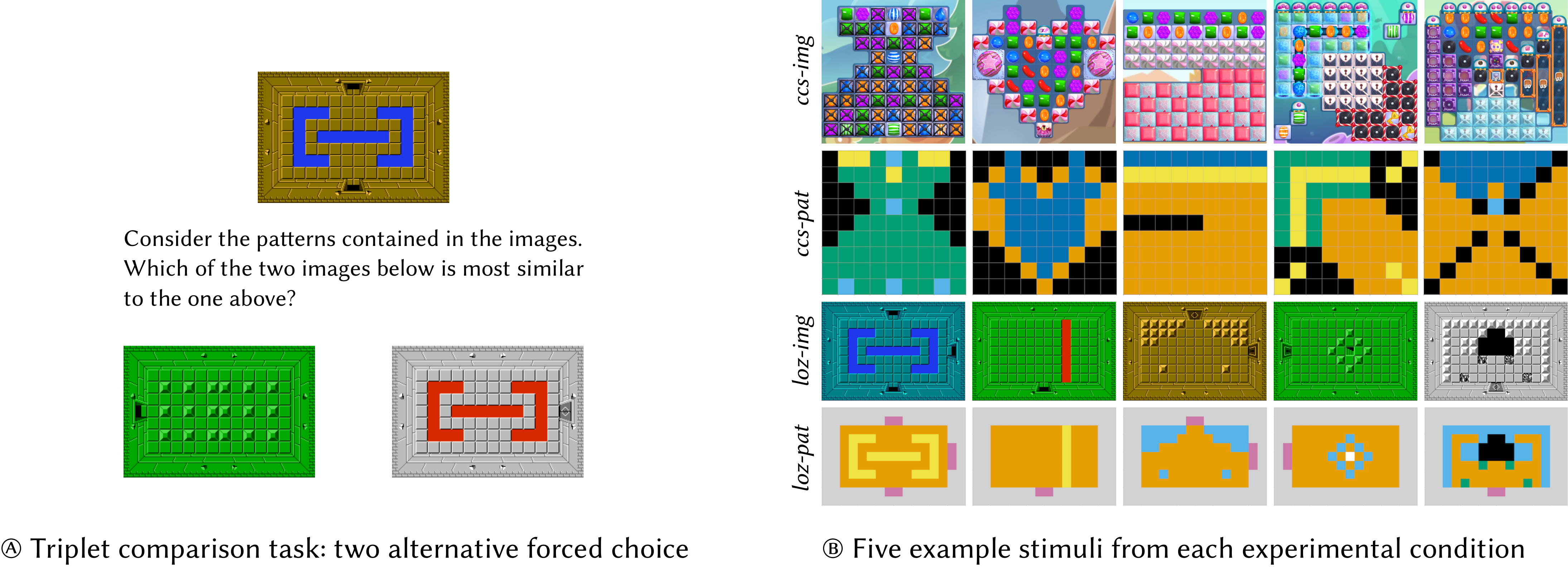}
  \caption{(A) Two alternative forced choice (2AFC) triplet comparison task and (B) five example stimuli for each experimental condition: two video game titles (\emph{ccs}: Candy Crush Saga; \emph{loz}: Legend of Zelda) in two representations (\emph{img}: level screenshots; \emph{pat}: abstract colour patterns). Each stimulus was randomly drawn from the respective subset identified through our three-stage selection procedure.}
  \Description{The triplet comparison task shows a reference stimulus A at the top and two stimuli B and C below as options to be selected by a participant. The instructions in the middle read: “Consider the patterns contained in the images. Which of the two images below is most similar to the one above?” The stimuli show examples of the tile-based video game levels used in our study. Candy Crush Saga levels are composed within a square grid (9 x 9 tiles) but often take horizontally or diagonally symmetric shapes (e.g. heart shape, diagonal crosses). Tile types include candies of various colours, which can be stuck in a transparent block of jelly, or be overlaid by a liquorice net. Legend of Zelda levels are all horizontally rectangular (16 x 11 tiles), showing a top-down view of a single room. Every room has a central floor area (12 x 7 tiles) with varying content (water, lava, enemies, blocks, downward-staircases), and is surrounded by 2-tile thick walls in all directions which may contain doors in their centre.}
  \label{fig:teaser}
\end{teaserfigure}

\received{14 September 2023}
\received[revised]{12 December 2023}
\received[accepted]{19 January 2024}

\maketitle

\section{Introduction}
\label{sec:intro}

For video games to be enjoyable, game designers must anticipate how their players will perceive, and consequently experience and react to, elements in the game. This however proves challenging when considering elements that only unfold dynamically at runtime, such as procedurally generated content or the behaviour of non-player characters (NPCs). To constrain such processes and meet players' expectations, designers can endow them with computational metrics that approximate player perception, experience, and behaviour \citep{yannakakis2011experience, canossa2015towards, guckelsberger2017predicting}.

Here, we focus on an important family of such metrics to assess players’ perception of visual similarity. Metrics of similarity are an integral part of many game AI applications such as procedural content generation (PCG), which we use as a motivating case. For example, procedural content generation via machine learning (PCGML) \citep{summerville2018procedural} approaches rely on similarity metrics to generate artefacts such as game levels that resemble existing samples, or that are sufficiently distinct from previously played levels. In contrast to this runtime use-case, similarity metrics have also been used in design-time tools, e.g.~to determine and modify the expressive range of content that a generator with a specific configuration can produce \citep{smith2010analyzing, summerville2018expanding, cook2021danesh}.

There exist many similarity metrics to choose from, ranging from general-purpose ones to data-driven approaches to expert measures custom-tailored to a scenario. Typically, game designers and researchers select a similarity metric based on conventions, personal preferences, basic assumptions, or computational properties.

However, in both online and design-time PCG practice, it is unclear if the generated content is actually perceived as similar by players, i.e. how well the metric works as a surrogate of players’ perception \citep{volz2020capturing,withington2023right}. If the selected metric is misaligned, the consequences can be detrimental to how a game is experienced, as exemplified by the \emph{Thousand Bowls of Oatmeal Problem}, a term coined by Kate Compton to describe the \enquote{common antipattern of generating a set of artefacts which are technically distinct to the computer, but perceived by humans as uniform} \citep{compton2015casual}. Given its strong resonance with game development reality, it has quickly become one of the best-known idioms in the PCG community \citep{rabii2023oatmeal}. A classic example of this phenomenon is the lack of visual variety in the 18 quintillion possible planets of \emph{No Man’s Sky}, which are mathematically, but not perceptually, unique~\citep{nomanssky_oatmeal}. In this context, perceptual uniqueness has been promoted as the \enquote{real metric} required. Its characterisation as a \enquote{darn tough} one, highlights the importance of the development and identification of visual similarity metrics that approximate human perception as a core research challenge in games.

We hold that there exists no empirical data in the context of games to adequately support designers and researchers in selecting the most appropriate similarity metric. At the same time, there exists a wide range of metrics to choose from, including general-purpose metrics with no psychometric claims, custom-made metrics from game AI and PCG, and models from computer vision (CV) research. While some hold that writing metrics that ought to approximate human perception in video games \enquote{is a difficult skill that requires a deep understanding of the application domain}~\citep{rabii2023oatmeal}, it is presently unclear whether such domain-specific metrics truly perform best. Image embedding models are of particular interest here. While CV has long been active in developing surrogate models for the human similarity judgement for very specific domains \citep[e.g.][]{wills2009toward, piovarvci2018perception, lagunas2019similarity, shi2021low}, we recently saw a surge in the publication of more generic models \citep[e.g.][]{radford2021learninga, fu2023dreamsim}. This goes against the above claim for the necessity of domain knowledge. Moreover, one may assume that such CV models cannot approximate the human similarity judgement well on the synthetic and highly stylised levels of especially non-realistic video games, since they are trained on, and optimised primarily for natural images. But we do not know this for a fact. Moreover, metric development at present is based on designer intuition, but we are in the dark about which visual features of game levels really determine players’ similarity judgement.
This is of high significance for the game industry and research, as the development of custom metrics is time-intensive and hence costly. More generally, choosing a sub-optimal metric could result in bad guidance at design time or unsatisfying player experiences when used in-game.

With this paper, we build a better understanding of the human similarity judgement and its alignment with existing metrics for a specific sensory modality in a well-constrained but very popular non-realistic game genre. We focus on people's perception of visual similarity of tile-based video game levels, and address two research questions: (RQ1) Which existing metrics approximate the human similarity perception of grid-based video game levels best? And, (RQ2) what are the dimensions of this space that govern players’ similarity perception? 
The second question serves as a direct response to the first, in that its answer can serve as a stepping stone to inform the development of better domain-specific metrics. Moreover, the gained insights can teach designers how players perceive their assets, e.g. to inform a more intuitive creative process at design time even if not relying on computational support tools.

We investigate these questions through two empirical studies. We collect data on the human similarity perception in a 2 x 2 factorial study, covering two very different titles (Candy Crush Saga; Legend of Zelda) in two visual representations (level screenshots; abstract colour patterns). In a mixed design, participants compared the similarity of level triplets for subsets of each factor combination.
In each triplet comparison task, they are presented with a reference stimulus and choose the most similar stimulus from two options. Choices are forced, and participants cannot skip a task.
Using a variant of multi-dimensional scaling (MDS), we build domain-specific perceptual spaces, encoding similarity-relevant attributes for this specific scenario.
We compare a selection of PCG, general purpose and computer vision metrics against these perceptual spaces, thus contributing to RQ1. This is complemented by our second study, in which we asked focus groups \emph{with relevant experience} to gather interpretations of the dimensions underlying these perceptual spaces, supporting RQ2 and thus fostering designer insights and the future development of better metrics. 

Our contributions are threefold:
\begin{enumerate}
    \item A quantitative study comparing similarity judgements from 456 participants against a total of 12 configurations of 7 existing metrics.
    \item A qualitative interpretation study, in which four focus groups \rev{with experience relevant to video game design, development, and research} (N=4x2) provide their interpretations of the dimensions underlying the human similarity judgements in this domain.
    \item A public dataset of human similarity judgements in tile-based video game levels and implementation of the comparison test suite.\footnote{\url{https://github.com/sebastianberns/similarity-estimation-chi24}}
\end{enumerate}
We moreover critically reflect on the requirements of each group of metrics and provide recommendations for scenarios when not the best but a runner-up metric might be preferred.

Similar to \citet{rabii2023oatmeal}, we thus set out to put intuitions and internalised knowledge of game researchers and practitioners to the test in the hope of strengthening applications and inspiring new research. Our findings from (1) serve the game development and research communities by informing recommendations for which existing metrics should be preferred in different scenarios, resting on a strong empirical basis. Moreover, our findings from (2) inform the future development of more human-aligned similarity metrics in the video games domain. The publicly available code and data in (3) facilitate the evaluation of additional existing and newly developed similarity metrics, thus enabling game developers and researchers to easily build on our work. Focusing on levels of tile-based video games, we relate to a particularly big and popular game genre, and arguably the most common type of procedurally generated content. While using PCG as a prominent application domain to motivate this research, an evaluation of these metrics in concrete PCG algorithms is out of scope. Instead, we focus on the comparison of existing metrics for approximating the human similarity perception in a way that is agnostic w.r.t. the specific game AI application.

We next provide a brief overview of related work in \cref{sec:related_work}, followed by background information on the similarity metrics considered in our study, and an introduction to perceptual embeddings in \cref{sec:background}. We report on our two studies in \cref{sec:study1} and \cref{sec:study2}, complemented with a discussion of our findings and study limitations in \cref{sec:discussion} and \cref{sec:limitations}. The paper closes with conclusions of our findings and an outlook on future work.

\section{Related Work}
\label{sec:related_work}
Studies on the human perception of similarity are at the core of psychophysics. They cover a vast space of stimuli, from more basic stimuli such as sound or colour to complex ones such as motion or 3D models. To the best of our knowledge, there exist no empirical studies on human similarity judgement and its comparison to surrogate metrics in the domain of video games. A taxonomy of game evaluation metrics put forward by Volz \citep[][Appendix A]{volz2019phd} suggests that very few of such metrics in games draw on insights into human perception, and of those few, none measure similarity. 
Previous related work on the alignment of computational metrics with human perception~\citep{marino2015empirical, summerville2017understanding} focuses on human perception of fun, difficulty, and aesthetics within individual levels. Arguably the closest predecessor to the present study, \citet{marino2015empirical} investigate whether a series of computational metrics used in PCG adequately capture player's perceptions of levels of Super Mario Bros. Amongst unrelated metrics, they calculate Compression Distance as a metric of structural dissimilarity between pairs of levels. Crucially though, they do not correlate it with the player’s perception, likely because the experimenters did not find significant differences in compression distance between the generated levels examined in the user study. In contrast, our present work focuses specifically on comparing similarity metrics to people's perception of similarity of game levels.

Our choices of data collection and analysis methods are well supported by related work. Human similarity ratings are typically collected by presenting participants with triplet matching tasks (\emph{two alternative forced choice}, 2AFC, \cref{fig:teaser})~\citep{wills2009toward, piovarvci2018perception, lagunas2019similarity, shi2021low, fu2023dreamsim}, which is the most robust judgement type~\citep{demiralp2014learning}. Collected triplet judgements are then converted into a perceptual space through multi-dimensional scaling (MDS) or related ordination techniques (see \cref{sec:background:embedding} for details). The resulting representation is often used to label and thus identify the dimensions underlying the human similarity judgement for the stimuli in question. Crucially, this is where most existing analyses stop; we adopt this common methodology for our study, but take it further by comparing the perceptual spaces derived from human judgement against those produced by computational metrics.

As motivated earlier (\cref{sec:intro}), an intriguing question is how well state-of-the-art computer vision metrics, which were not developed specifically for use in games, can compete with more conventional or custom-made metrics already adopted in games. Crucially though, stimuli in studies on image similarity more generally, e.g.~photographs \citep{rogowitz1998perceptual}, are arguably far removed from the imagery that players experience in the tile-based video games under consideration. We include DreamSim~\citep{fu2023dreamsim} here both as a recent example to frame and compare our study setup against, as well as a metric in our study (\cref{sec:study1}). For the development of DreamSim, \citet{fu2023dreamsim} have curated a dataset of human judgements over pairs of synthetic images, following the same 2AFC triplet judgement task method described above and employed in our work. Crucially, their image triplets were iteratively selected for maximum participant agreement, effectively optimising for an easily solvable binary decision task. In contrast, we take into account participant disagreement and thus gather richer relational information between stimuli. Their work focuses on synthesised natural images and thus compares conventional CV metrics and state-of-the-art learned, i.e. data-driven, embeddings. We instead focus on metrics relevant to video game development and research but overlap with their work in comparing CLIP \citep{radford2021learninga} as a popular image embedding. They finally use their dataset to fine-tune an ensemble model for measuring image similarity, which we leave for future work.

\section{Background}
\label{sec:background}

We first introduce how similarity is measured in the context of video games and present the specific metrics compared in our study. We then explain how to construct perceptual embedding spaces from triplet judgement data.

\begin{table*}[t]
\caption{Selection of \rev{image embeddings, metrics and} measures (with optional configurations) \rev{compared in this work}. Note that the \rev{image embeddings and} measures require additional transformations to be used as similarity metrics (\cref{sec:background:metrics}).}
\label{tab:measures}
\centering

\begin{tabularx}{\textwidth}{lllll}
\toprule
Name & Group & Input & Output & Description \\
\midrule
CLIP \citep{radford2021learninga} & CV & Image & Vector & \makecell[l]{Image embedding trained on a huge dataset of image-text pairs\\scraped from the internet.} \\
\midrule
DreamSim \citep{fu2023dreamsim} & CV & Image & Vector & \makecell[l]{Image embedding fine-tuned on human similarity judgements\\(two alternative forced choice).} \\
\midrule
\makecell[l]{Normalised Compres-\\sion Distance \citep{li2004similarity}} & General & Tiles & Scalar & \makecell[l]{Using a compression algorithm (gzip), compares the joint compression\\length of two levels to their individual compression lengths.} \\
\midrule
Hamming Distance & General & Tiles & Scalar & Fraction of tiles that exactly match across two levels. \\
\midrule
Tile Frequencies \citep{summerville2017understanding} & PCG & Tiles & Distribution & Relative frequencies of tile types appearing in a level. \\
\midrule
Tile Patterns \citep{lucas2019tile} & PCG & Tiles & Distribution & \makecell[l]{Relative frequencies of tile patterns appearing in a level.\\Configurations: size of patterns ($2\times2$, $3\times3$, $4\times4$).} \\
\midrule
Symmetry \citep{volz2020capturing} & PCG & Tiles & Scalar & \makecell[l]{Fraction of tiles that match when mirroring half of a level across a cor-\\responding axis (e.g. vertical symmetry: left and right halves compared\\across the centre). Configurations: axis of symmetry (Horizontal, Vertical,\\Diagonal Forward, Diagonal Backward).} \\
\bottomrule
\end{tabularx}
\end{table*}

\subsection{Similarity Metrics for Video Games}
\label{sec:background:metrics}

To calculate video game level similarity, game developers and researchers leverage methods from three different groups of measures and distances: 1) artificial neural network-based image embeddings trained on datasets of natural images for computer vision (CV), 2) domain-agnostic, general-purpose distance metrics (General), and 3) manually-designed measures based on expert knowledge, from the PCG literature (PCG). In \cref{tab:measures}, we list and describe all embeddings, distances, and measures used as metrics for comparison in this study. Our focus in this selection lies on measures specifically used in video games-related research and the game industry. In the following, we provide further detail about our choice of metrics.

We define a \emph{measure} as a method to quantify the qualities of a video game level and a \emph{metric} as the comparison of such qualities between two levels.\footnote{Note that we do not follow the stricter mathematical definition of a “metric” here, but instead use the term more colloquially as a way of more easily differentiating the methods that quantify qualities from those that compare them.} To build a working similarity metric, embeddings, distances, and measures need to be transformed and compared. We outline here how this applies to the aforementioned groups and our selection.

In computer vision (CV), it is common to use the embedding spaces of artificial neural networks to compare the perceptual similarity of images \citep{zhang2018unreasonable}. Most recent embedding models (e.g. CLIP) have been specifically designed for the evaluation of two inputs via cosine similarity \citep{radford2021learninga}. We chose two state-of-the-art image embedding models: CLIP (ViT-L/14@336px) for its ubiquitous use and DreamSim (ensemble) for its specific alignment with human perception. Both take as input one square colour image (either a level screenshot or the corresponding colour pattern; see \cref{sec:study1:materials}) and yield its corresponding embedding vector. To evaluate the similarity of any pair of images, we calculate the cosine similarity between their embedding vectors.

While little PCG research focuses on similarity estimation specifically, many works propose or use some measure to evaluate generative systems and their output. For example, in expressive range analysis \citep{smith2010analyzing} or to drive quality diversity search in video game asset production \citep{fontaine2021illuminating}. Researchers draw from expert knowledge to design specialised measures that capture relevant qualities. In contrast to CV embedding models, PCG measures always take as input a tile-based representation of a level (independent of experimental condition), where individual tile types are encoded as ASCII characters. Tile Frequencies is a popular baseline measure to characterise tile-based levels \citep{summerville2017understanding}. While it disregards the location of tiles and thus does not fully capture the composition of a level, the discrepancy of different tile types appearing in two levels might be enough to approximate the overall visual similarity between the two levels. This simple idea has been extended to larger Tile Patterns \citep{lucas2019tile}. While Tile Frequencies only consider individual tiles ($1\times1$ patterns), Tile Patterns can be configured to calculate the occurrences of any $N\times M$ pattern in a level. Both Tile Frequencies and Tile Patterns take as input the tile-based representation of one level and yield the probability distribution over the tiles or patterns that appear in the level. We calculate the similarity between two levels by first calculating the Jensen-Shannon distance between the two tile or pattern distributions and then converting their distance into similarity by subtracting it from 1. We further included symmetry measures because research on patterns in Candy Crush Saga has shown that symmetric generated levels are considered more similar to original game levels by human expert judges \citep{volz2020capturing}. While symmetry by itself is probably not sufficient to fully describe level similarity, we hypothesise that it might be an important factor in the human perception of tile-based video game levels. Symmetry measures take as input one level in a tile-based representation and yield a scalar output that quantifies the level’s symmetry on a given axis (horizontal, vertical, or either forward or backward diagonal). Two levels are compared in terms of their similarity by calculating the absolute difference between their symmetry scores.

As general and domain-agnostic metrics, we selected Hamming and Normalised Compression Distance (NCD). While these have been applied to PCG \citep{rodrigueztorrado2020bootstrapping, edwards2021searchbased}, they have not been specifically developed for video game applications. Instead, they stem from informa\-tion-theoretic approaches to measuring distances between strings of text. Hamming Distance provides a simple baseline, is easily interpretable and finds many applications in video game research in its more general form as \emph{edit distance} \citep{alvarez2018assessing,graham2023level}. NCD has been used as a metric for the structural similarity of video game levels, as it encodes both tile frequencies and positions \citep{marino2015empirical,shaker2012evolving}. Both general metrics take as input two levels in the tile-based representation and yield the distance between them, which is converted into their similarity by subtracting the distance from 1.

\subsection{Perceptual Embedding Spaces}
\label{sec:background:embedding}

Our work aims to compare various similarity metrics to the judgements collected from our study participants. To facilitate this, we apply the conventional methodology of constructing a \emph{perceptual space} from the triplet judgements that embeds all stimuli in a Euclidean space where distances correspond to the perceived relations between triplets \citep{demiralp2014learning, piovarvci2016interaction, piovarvci2018perception}. We thus understand the distance in the Euclidean space to be the inverse of perceived similarity: the more similar two stimuli are, the closer they will be positioned to each other in the embedding.

More formally, in an exemplary triplet judgement task, let $A$ be the reference stimulus, and $B$ and $C$ be the two options participants can choose from (\cref{fig:teaser}). Suppose a participant decides that the reference $A$ is more similar to option $B$ than it is to option $C$. We can describe this relation as $d(A,B) < d(A,C)$, where $d$ is a distance metric in Euclidean space. Let us call this a \emph{paired comparison} of the given triplet $A,B,C$. The embedding space is built by finding the vectors corresponding to all stimuli $\Vec{a}, \Vec{b}, \Vec{c}$, such that $\lVert \Vec{a} - \Vec{b} \rVert < \lVert \Vec{a} - \Vec{c} \rVert$. Naturally, this relationship should hold for all collected triplet judgements, thus creating a set of constraints on the vectors. The construction of the perceptual embedding is conventionally formulated as a constrained optimisation problem.

A common method to obtain such an embedding is multi-dimen\-sional scaling (MDS). A loss function (called \emph{strain}) quantifies how well the embedding satisfies all constraints. Several versions of MDS exist, most notably metric and non-metric algorithms. However, most require a target distance matrix in which the pairwise similarities between stimuli are expressed as numerical distances. This is difficult to obtain from our study data, in particular, because not all participants judged every triplet. Generalised non-metric multi-dimensional scaling (GNMDS) \citep{agarwal2007gnmds} instead reformulate the loss function to primarily depend on information from the paired comparisons. Additional slack variables account for unsatisfied constraints. The optimisation objective aims to minimise the amount of slack. Yet, when there is high disagreement in the data between judgements from individual participants, it becomes difficult to satisfy all constraints at once. This results in large amounts of remaining slack.

In our work, we employ t-distributed stochastic triplet embedding (t-STE)~\citep{vandermaaten2012stochastic}, which responds better to the naturally occurring noise in the judgement data by not trying to satisfy constraints that contradict the consensus. t-STE can thus deal best with two important characteristics of our collected judgement data: 1) missing data due to participants only judging a subset of triplets, and 2) high disagreement between individual participants due to the difficulty of the triplet judgement task.

\section{Study 1: Human vs. Computational Similarity Evaluation}
\label{sec:study1}

To compare the human evaluation of similarity with computational metrics, we collect data on people's evaluation of similarity in tile-based video games. To this end, we employ a full 2x2 factorial design with the first factor defining the video game \emph{Title} (\emph{ccs}: Candy Crush Saga; \emph{loz}: Legend of Zelda) and the second the visual \emph{Representation} of levels (\emph{img}: level screenshots; \emph{pat}: an abstract colour tile pattern of the level sprite layout). This yields a total of four experimental conditions: \emph{ccs-img}, \emph{ccs-pat}, \emph{loz-img}, \emph{loz-pat}. 
We choose two \emph{Representations} to cover different scenarios relevant to the application of similarity metrics in video games and PCG. The \emph{img} representation provides direct insight into how people assess the similarity between levels as they appear in the given \emph{Titles}. Through this, we aim to inform the selection of similarity metrics for application in these and other closely related video games. 
With the \emph{pat} representation, we focus on more abstract pattern representations of level layouts as they are used in the level design process and by many PCG algorithms. Our goal is to provide practical recommendations for game designers, developers and researchers for the application of similarity metrics at design time and in conjunction with PCG and PCGML approaches.
We are interested in answering RQ1 individually for both of these scenarios (\emph{Which existing metrics approximate the human similarity perception of grid-based video game levels best?}).
Note that there is an important difference in the design of the similarity metrics. CV-based metrics (CLIP and DreamSim) take image input and can be applied to any image. In \emph{img} conditions, they will be given level screenshots, whereas, in \emph{pat} conditions, they will receive the colour patterns. In contrast, all other metrics receive levels in their tile-based representation and are given the same information in all conditions.

With the stimuli in each condition, we prepared a collection of triplet comparison tasks as two alternative forced choice (2AFC) questions. Given a reference stimulus, participants are asked to make a forced choice between two stimuli, selecting the option most similar to the reference. This design was shown to be the most robust data collection method and has been recommended for assessing perceptual similarities (\emph{triplet ranking with matching})~\citep{demiralp2014learning}. To prevent participant fatigue but still assess a high number of stimuli, we employ a mixed design where each participant judges a subset of triplets from each condition. The study was approved by \rev{the Queen Mary Ethics of Research Committee}.\footnote{\rev{Reference number:
QMERC20.565.DSEECS23.030}}

\begin{table*}[t]
\caption{Self-reported experience with tile-based video games of participants in \textcolor{gamesexp-blue}{study 1 (blue)} and \textcolor{gamesexp-red}{study 2 (red)}. Participants selected one option in each row, and percentages in each row add up to 100\,\%.}
\label{tab:gamesexp}
\begin{tabular}{l|llll}
 &
  \begin{tabular}[c]{@{}l@{}}I do not know this \\ type of game\end{tabular} &
  \begin{tabular}[c]{@{}l@{}}I have heard of \\ this type of game\end{tabular} &
  \begin{tabular}[c]{@{}l@{}}I have played this \\ type of game\end{tabular} &
  \begin{tabular}[c]{@{}l@{}}I regularly play this \\ type of game\end{tabular} \\ \hline
\begin{tabular}[c]{@{}l@{}}Tile-matching games (like\\Candy Crush or Bejeweled)\end{tabular} & \textcolor{gamesexp-blue}{1.80\%} \textcolor{gamesexp-red}{—} & \textcolor{gamesexp-blue}{14.7\%} \textcolor{gamesexp-red}{25\%} & \textcolor{gamesexp-blue}{59.1\%} \textcolor{gamesexp-red}{75\%} & \textcolor{gamesexp-blue}{24.4\%} \textcolor{gamesexp-red}{—} \\
Pacman or Ms Pacman & \textcolor{gamesexp-blue}{2.90\%} \textcolor{gamesexp-red}{—} & \textcolor{gamesexp-blue}{16.5\%} \textcolor{gamesexp-red}{—} & \textcolor{gamesexp-blue}{73.6\%} \textcolor{gamesexp-red}{100\%} & \textcolor{gamesexp-blue}{7\%} \textcolor{gamesexp-red}{—}  \\
\begin{tabular}[c]{@{}l@{}}Retro dungeon crawlers\\ (like Legend of Zelda)\end{tabular} & \textcolor{gamesexp-blue}{29\%} \textcolor{gamesexp-red}{—} & \textcolor{gamesexp-blue}{42.0\%} \textcolor{gamesexp-red}{37.5\%} & \textcolor{gamesexp-blue}{23.5\%} \textcolor{gamesexp-red}{37.5\%} & \textcolor{gamesexp-blue}{5.50\%} \textcolor{gamesexp-red}{25\%}  \\
Sokoban & \textcolor{gamesexp-blue}{75.8\%} \textcolor{gamesexp-red}{62.5\%} & \textcolor{gamesexp-blue}{15.6\%} \textcolor{gamesexp-red}{12.5\%} & \textcolor{gamesexp-blue}{7.70\%} \textcolor{gamesexp-red}{25\%} & \textcolor{gamesexp-blue}{0.9\%} \textcolor{gamesexp-red}{—}  \\
\begin{tabular}[c]{@{}l@{}}Bomberman, Dyna Blaster,\\or similar\end{tabular} & \textcolor{gamesexp-blue}{48.1\%} \textcolor{gamesexp-red}{12.5\%} & \textcolor{gamesexp-blue}{20.7\%} \textcolor{gamesexp-red}{—} & \textcolor{gamesexp-blue}{29.9\%} \textcolor{gamesexp-red}{87.5\%} & \textcolor{gamesexp-blue}{1.3\%} \textcolor{gamesexp-red}{—}  \\ \hline
\end{tabular}
\end{table*}

\subsection{Materials}
\label{sec:study1:materials}

As stimuli, we first select a subset of level images from both video games (\emph{Title}).
Video game levels in the \emph{img} representation include some decorative elements, e.g. different colour sprites for the same game objects in \emph{loz} and certain game objects, like candies, being represented by different sprites in \emph{ccs}. We hypothesise that the \emph{img} representation, essentially content shown in-game, evokes gameplay associations in the participants and obfuscates some similarity-relevant visual patterns.

To test this hypothesis, we leverage an abstract colour pattern representation (\emph{pat}) for each \emph{Title} that relies on existing mappings from level object to colour tile (\citet{summerville2016vglc} for \emph{loz} and \citet{volz2020capturing} for \emph{ccs}). The purpose of the \emph{pat} representation is to remove potentially distracting gameplay associations and emphasise the similarity-relevant characteristics of levels, e.g. shapes and patterns. These types of colour tile patterns are commonly used in PCG in research \citep{summerville2016vglc, summerville2018procedural, sarkar2022tile2tile, bhaumik2023lode}, as well as in practice \citep{staalberg2018bad, grinblat2010caves}. 
\rev{In the following we describe how the conversion from \emph{img} to \emph{pat} representation is performed and how it differs between the two \emph{Titles}. However, to represent level objects, we apply the same colour-blindness-safe colour palette \citep{wong2011points} to converted level representations from both \emph{Titles}}.
For Legend of Zelda (\emph{loz}), the colour tile mapping defined in the VGLC \citep{summerville2016vglc} is straightforward, as it simply maps level elements with different functionality to distinct colour tiles (e.g. walls are different from floors are different from enemies are different from doors, and thus assigned different colours). In this abstraction, simplifications are limited to subsuming all enemies into a single colour tile and ignoring the different colour palettes of the various dungeon rooms. 
The colour tile mapping we use for Candy Crush Saga (\emph{ccs}) is informed by \citet{volz2020capturing} and was devised in collaboration with the game’s creator, King. Instead of a direct one-to-one mapping from each level object to a colour tile, this representation subsumes several level objects with similar in-game behaviour. For example, objects that look visually distinct (e.g. frosting and chocolate objects) but perform similar game functions (\emph{blockers} impede moves, making gameplay more difficult) are mapped to the same colour tile. This is done for several functional level objects, such as \emph{blockers}, \emph{candies}, \emph{power pieces}, and \emph{locks}. 
Using the above \emph{img} to \emph{pat} mappings, we represent every \emph{Title} in each \emph{Representation}. For details on the specific conversion procedures, please refer to our implementation in the supplementary material.

As datasets, we obtain \emph{loz} levels from an open-source corpus of video game levels~\citep{summerville2016vglc} and scrape \emph{ccs} level screenshots from a fan wiki\footnote{\url{https://candycrush.fandom.com}}.
Since our datasets (\emph{ccs}: 2,792 levels, \emph{loz}: 225 levels) were too large for a triplet comparison study, we selected a subset of stimuli informed by the expected amount of participants, a minimum of five comparisons per triplet, and a maximum amount of 100 comparisons per participant. To select a subset representative of the overall variety in levels, we employed a three-stage selection pipeline. We first obtained image embeddings from an artificial neural network (CLIP ViT-14/L@336px \citep{radford2021learninga}). While using a metric assessed in the same study introduces a bias, the bias is explicit and can be accounted for. We discuss this and alternatives we considered in the limitations \cref{sec:limitations}. We then reduced the dimensionality of the embeddings from 768 to 2 dimensions with t-SNE \citep{vandermaaten2008visualizing}, to make the subsequent sampling step feasible. In our implementation, t-SNE uses cosine similarity, which is the most appropriate to calculate distances between CLIP embeddings. The origin of biases is thus limited to the choice of embedding model. Finally, we used conditioned Latin Hypercube Sampling (cLHS)~\citep{minasny2006conditioned} to find a subset for which items are maximally distant from each other in the low-dimensional embedding space. This is to ensure that 1) the samples cover a large part of the space of possible levels and 2) that we do not inadvertently draw conclusions from a non-representative subset of levels. To mitigate the influence of different tile colours in Legend of Zelda, we select levels based on their greyscale versions. We selected 17 stimuli for each of the four experimental conditions, yielding $\binom{17}{1} \times \binom{16}{2} = 2040$ triplets per condition, and $8160$ triplet comparisons overall. \Cref{fig:teaser} shows a random selection of five levels from all subsets, each corresponding to one condition. 

Participants were asked optional demographic questions about their self-described gender and age, and their experience with tile-based video games. The surveys were implemented in Qualtrics. Given a list of stimuli, we compute all triplet combinations and generate individual surveys for all conditions for upload to Qualtrics.

\subsection{Participants}
\label{sec:study1:participants}

We recruited 460 participants from Prolific to complete a 15-minute survey paid at the equivalent of an hourly rate of £10. Funding was provided by modl.ai. We excluded four participants who did not complete the full survey and proceeded with the data from the remaining 456 participants. Out of these, 53.51\,\% reported their gender as female, 43.64\,\% as male, 1.75\,\% identified as non-binary or third-gendered, none chose to self-describe, 0.44\,\% preferred not to respond, 0.44\,\% left the question unanswered, and 0.22\,\% abandoned the survey before seeing the question. The median reported age is 28. Our sample is thus considerably more representative w.r.t. identified gender than common in studies related to video games. We summarise their self-reported experience with tile-based video games in \cref{tab:gamesexp}.

\subsection{Procedure}
\label{sec:study1:procedure}

We informed our final study procedure based on a pilot, involving seven stimuli in each condition. The goal of this pilot was to test the survey setup and identify average response times, suitability of validation questions, and baseline disagreement ratios on individual triplets. It was completed by 22 trusted participants from the authors’ respective industry and academic institutions. 

Our study follows the conventional methodology for collecting human similarity ratings with \emph{two alternative forced choice} (2AFC) questions, one of the oldest methods of psychophysics \citep{fechner1860elemente}. We interchangeably refer to this as triplet comparisons. Given a reference stimulus, participants are asked to make a forced choice between two stimuli, selecting the option most similar to the reference. For our study, out of the $8160$ total triplets (\cref{sec:study1:materials}), every participant was assigned a random subset of $25$ from each of the four experimental conditions. In addition to these $100$ triplet comparisons, participants were asked to judge three additional triplets as validation questions in each condition. The order between and within conditions was randomised for each participant, and colour patterns were shown before level images to not prime participants’ perceptions. Participants provided informed consent at the start of the survey and answered optional questions on demographics and game experience after judging all triplets.

\subsection{Data Analysis}
\label{sec:study1:analysis}

To understand how the computational metrics correlate with our data on the human perception of similarity, we perform two complementary quantitative data analyses. 
First, we quantify how well the computational metrics can approximate the similarity matrices derived from our participant data. For this, we construct a perceptual space for each condition which embeds the stimuli in a low-dimensional Euclidean space. Second, we conduct pairwise comparisons between the judgements of individual human participants and the different computational metrics in an inter-rater agreement analysis. 
In addition, we provide a qualitative analysis of the features underlying the human similarity judgements in our second study (\cref{sec:study2}). All analyses are performed separately for each experimental condition.

\subsubsection{Perceptual Embedding of Tile-Based Level Similarity}
\label{sec:analysis:embedding}

To determine the overall relationships between stimuli in terms of similarity, aggregated over all human responses, we construct a perceptual space from the collected triplet judgements, i.e.~an embedding of stimuli in Euclidean space (here also called \emph{perceptual embedding} or \emph{embedding space}). Participants were asked for their subjective perception of similarity. Choices were forced, and participants did not have the option to skip a judgement task. It is natural that the triplet data is noisy and reflects some disagreement. Yet, this provides important information about the similarity-relations of stimuli and introduces constraints that need to be taken into account. For example, if many participants agree that reference stimulus $A$ is more similar to stimulus option $B$ than the other option $C$, in the embedding space $A$ needs to be positioned closer to $B$ than $C$. A perceptual embedding converts each individual piece of relationship information into an aggregated positional distance within the embedding while satisfying all constraints as best as possible. As noted in \cref{sec:related_work}, this inclusive approach also distinguishes our work from related work.

We chose the embedding algorithm t-distributed stochastic triplet embedding (t-STE)~\citep{vandermaaten2012stochastic} as it provides several advantages over conventional multi-dimensional scaling (MDS) methods, in particular, the handling of missing data and noisy data (for background see \cref{sec:background:embedding}). The former is necessary since not all participants judge all triplets, and the latter as our data shows a lot of disagreement between participants. Elbow plots (\cref{fig:elbows}) indicate that four dimensions can adequately encode the most relevant attributes across experimental conditions while providing a close-to-optimal fit for the triplet data.

To quantify the suitability of the embedding for subsequent comparisons with computational metrics, we analyse the robustness of the embedding to random initialisation over 10 runs with different random seeds. The goodness of fit to the raw data, number of required iterations, and number of constraints are almost identical across all runs. While the absolute positions of triplets in the embedding depend on the initialisation of the embedding and can differ significantly between random seeds, the variance of pairwise similarities between the embedded stimuli is much lower across all conditions (\emph{ccs-img}: 0.0349, \emph{ccs-pat}: 0.0464, \emph{loz-img}: 0.0389, \emph{loz-pat}: 0.0419; variance over 10 runs with random initialisation), indicating overall robustness of the resulting perceptual embeddings. For each condition, we select the embedding with the best fit to the data from these 10 candidates.

\subsubsection{Comparison of Similarity Matrices}
\label{sec:analysis:similarity}

\begin{figure*}[t]
  \centering
  \includegraphics[width=\linewidth]{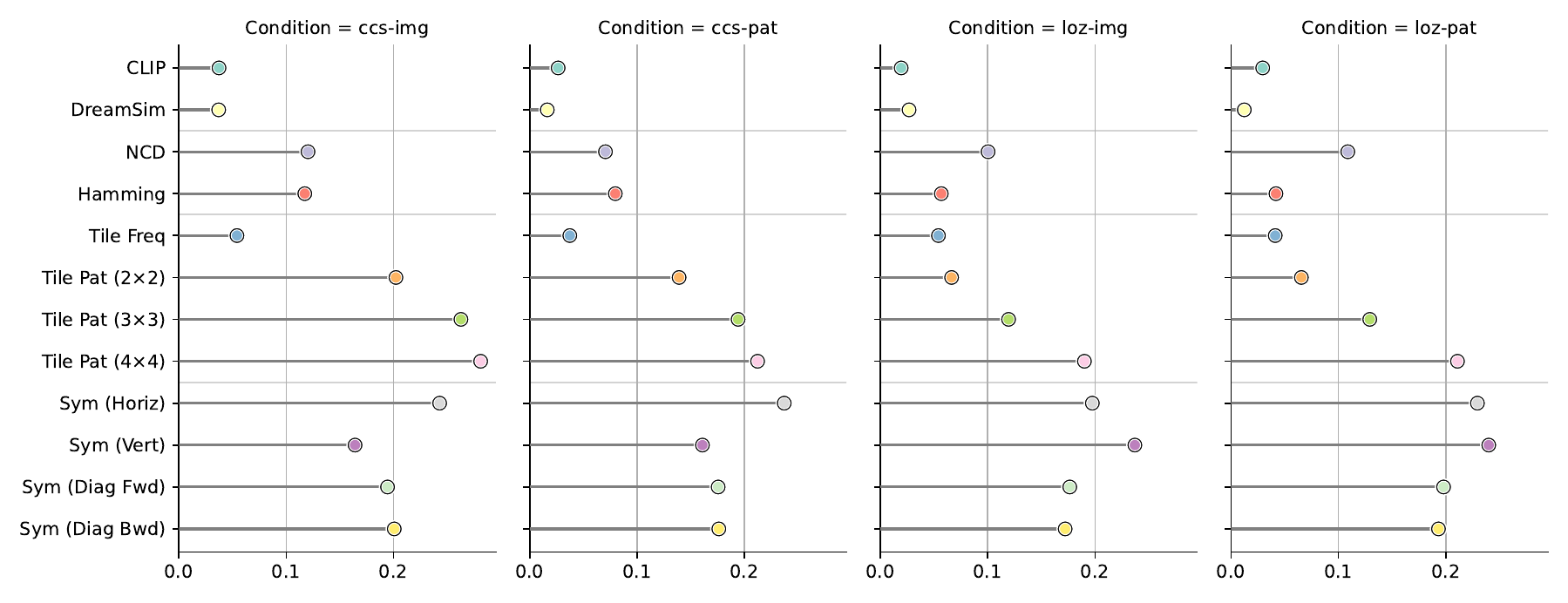}
  \caption{Mean squared errors (lower is better; horizontal axes) when comparing the pairwise similarity matrices of different candidate metrics (vertical axis) to those derived from the perceptual embeddings of the four experimental conditions (subplots).}
  \Description{Four scatter subplots, aligned next to each other, with results for all metrics in the four experimental conditions (from left to right: ccs-img, ccs-pat, loz-img, loz-pat).}
  \label{fig:study1:sim-mse}
\end{figure*}

To quantify the capabilities of the computational metrics to approximate the human similarity judgements, we calculate the error between similarity matrices derived from either source. The similarity matrix for human judgements is based on the previously described perceptual embeddings (\cref{sec:analysis:embedding}). We first compute the pairwise Euclidean distances between all stimuli in the embedding, then normalise them by the maximum distance, and finally convert normalised distances into similarities by subtracting them from 1. The similarity matrix for a computational metric is constructed from the pairwise similarity between stimuli computed by a given metric as outlined and motivated in~\cref{sec:background:metrics}. Two similarity matrices are compared by calculating the mean squared error. Results are summarised in \cref{sec:study1:results} and visualised in \cref{fig:study1:sim-mse}. 

This comparative analysis allows us to quantify a metric’s prediction error of the similarity-relation between two stimuli by comparing it to the ground-truth human perception. However, this can only be done by way of constructing a perceptual embedding space from the collected judgement data, which itself only approximates the judgement data. We supplement this first analysis with the following inter-rater agreement analysis, as it allows for a more direct comparison to the judgement data given by our participants, without requiring an intermediate approximation.

\subsubsection{Agreement between Participants and Metrics}
\label{sec:analysis:agreement}

We perform an inter-rater agreement analysis between human participants and computational metrics. Cohen’s kappa ($\kappa$) is calculated for pairs of one participant and one metric as the two raters. For this, we first find the triplets judged by a given participant and then determine the judgements of the metric in question on the same triplet comparison tasks. This allows us to perform a direct inter-rater agreement analysis. This process is repeated for each combination of participant and metric in each condition. We remind the reader that in each triplet comparison, a participant is presented with a reference stimulus A and chooses the most similar stimulus from two options B and C. In standard inter-rater agreement terminology, we thus deal with two raters each judging 25 items on a two-category nominal scale (stimulus option B or C). We use Cohen's kappa over agreement percentage, as it takes into account the possibility of chance agreements, which is particularly important when dealing with only two categories. As not every participant has judged all triplets, the statistics only reflect agreement on each participant’s random subset of 25 triplets from each condition. For each condition, we thus collect as many data points (kappas) as there are participants who completed this section of the survey. We further perform additional analyses of the inter-rater agreements, available in \cref{appendix:study1}, which support the results presented here. For ease of interpretation, however, we here focus on Cohen’s kappa.

This collection of inter-rater agreement statistics complements the initial comparison of similarity matrices, allowing for a direct comparison of a metric’s binary prediction to a given participant’s judgements. However, a binary choice between two stimuli options only gives a limited account of the complex similarity-relations between stimuli. In contrast, the initial comparison of similarity matrices can accommodate more fine-grained relations, expressed as distances in a Euclidean space.

\begin{figure*}[t]
     \centering
    \includegraphics[width=\textwidth]{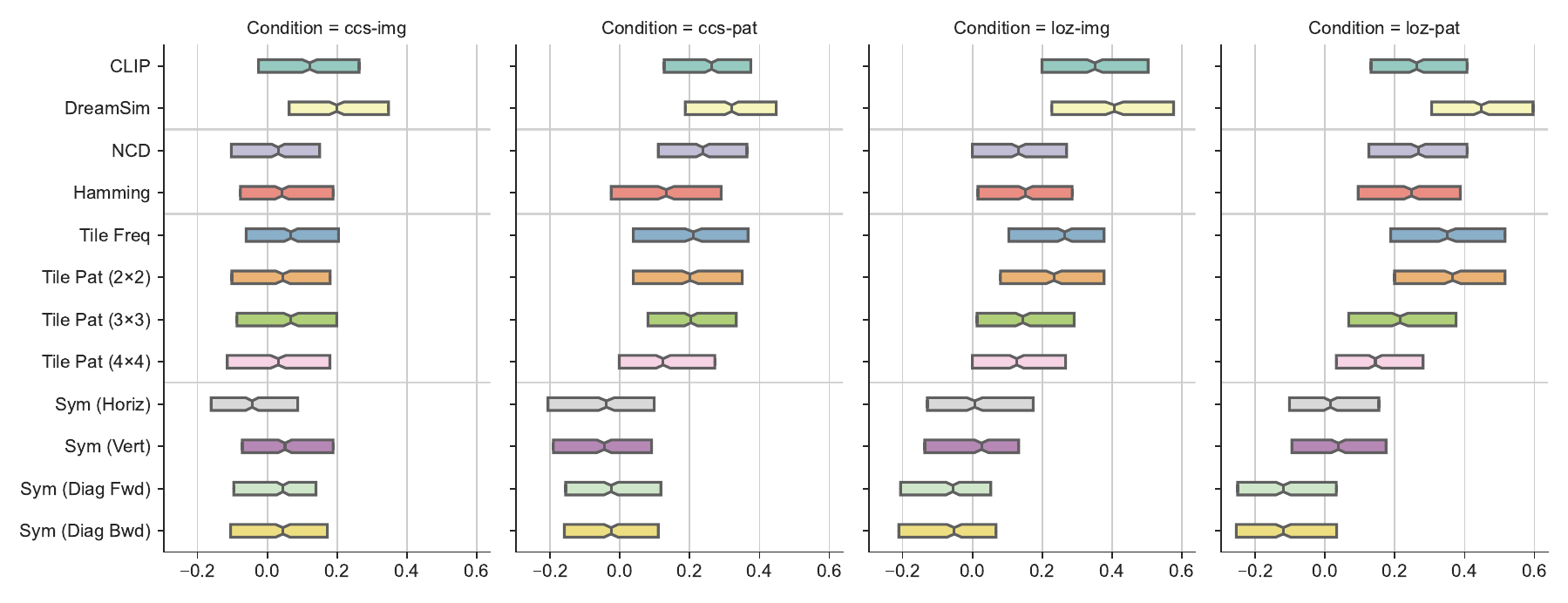}
    \caption{Cohen’s kappa (higher is better): inter-rater agreement between human participants and computational metrics over all experimental conditions (subplots). Summaries here show box plots with median values and the interquartile ranges. Full raincloud plots can be found in \cref{appendix:study1}.}
    \Description{Four subplots of summarising box plots for all metrics in the different experimental conditions (from left to right: ccs-img, ccs-pat, loz-img, loz-pat). The subplots share a vertical axis which lists all computational metrics.}
    \label{fig:agreement-boxplots}
\end{figure*}

\subsection{Results}
\label{sec:study1:results}

From the 456 participants, we collect a total of 11,400 judgements per condition, resulting in an average of 5.6 judgements per triplet comparison. We next present our results separately for each of the analysis steps outlined in \cref{sec:study1:analysis}. We give a summary of the results here, alongside visualisations. For full and exact results consult the supplementary material.

In our main analysis, we compared the pairwise similarity matrices of different candidate metrics to those derived from the perceptual embeddings. We report results as mean squared error, where a lower score indicates a better approximation of the human judgements by a computational metric (\cref{fig:study1:sim-mse}). Overall, the two CV metrics, CLIP and DreamSim, have the lowest errors across all experimental conditions. Looking at individual conditions, DreamSim exhibits the best approximation performance when using pattern-based representations (\emph{ccs-pat, loz-pat}). While CLIP has slightly lower error for image-based Legend of Zelda levels (\emph{loz-img}), both CV-based metrics are tied on image-based Candy Crush Saga levels (\emph{ccs-img}).
Tile frequencies are the overall third-best-performing approximate metric across all experimental conditions. In fourth place, the general-purpose metrics, Normalised Compression Distance (NCD) and Hamming Distance are tied in terms of overall error on Candy Crush Saga levels (\emph{ccs-img}, \emph{ccs-pat}). However, for Legend of Zelda levels in both representations (\emph{loz-img}, \emph{loz-pat}), Hamming Distance performs almost equally as well as Tile Frequencies. Tile Patterns in various configurations ($2\times2$, $3\times3$ and $4\times4$) are not good approximations for our collected human judgements. We observe that a larger pattern size leads to a higher error. We provide an explanation in the discussion (\cref{sec:discussion}). Similarly, Symmetry metrics in all configurations (horizontal, vertical, as well as diagonal forward and backward) yield comparatively high overall errors.

In a supporting inter-rater agreement analysis, we calculated the agreement between every pair of individual human participants and computational metric. A summarising box plot shows the median agreements and the interquartile ranges, where a higher score indicates higher agreement (\cref{fig:agreement-boxplots}; full raincloud plot available in \cref{appendix:study1}). As all metrics exhibit roughly similar interquartile ranges, we will focus our description on their median agreements with participant judgements. Out of all metrics, DreamSim shows the overall highest agreement. This is followed by CLIP which has the second-highest agreement in most conditions. The only exception is the pattern-based representation of Legend of Zelda levels (\emph{loz-pat}), where Tile Patterns ($2\times2$) and Tile Frequencies beat CLIP and share a close second-highest agreement. Tile Frequencies has the third-highest agreement for images of Legend of Zelda levels (\emph{loz-img}), again closely followed by Tile Patterns ($2\times2$). For the pattern-based representation of Candy Crush Saga levels (\emph{ccs-pat}), Normalised Compression Distance has the third-highest agreement, but only by a small margin when compared to Tile Frequencies as well as Tile Patterns ($2\times2$) and ($3\times3$). Third-highest agreement in Candy Crush Saga images (\emph{ccs-img}) is shared by Tile Frequencies and Tile Patterns ($3\times3$), though very closely followed by several other metrics.

We performed statistical significance testing \citep{vornhagen2020statistical} on the agreement between metrics and participant judgements (Cohen's kappa).
First, we test our basic assumption: (H1) there are significant differences in the performance of metrics in individual conditions. For this, we perform a one-way ANOVA separately for each condition. We further seek to evaluate two other hypotheses: (H2) DreamSim, from the CV group, performs better than Tile Frequencies, the next-best metric from a different group, i.e. PCG expert metrics; (H3) metrics have higher agreement with participant judgements of the pattern-based representation of levels than with judgements of level images. H2 is tested with a paired student’s t-test of the two related samples within individual conditions: participant agreement with DreamSim and with Tile Frequencies. H3 is tested with a separate independent student’s t-test of each metric between individual conditions. As H3 entails multiple comparisons, we correct p-values with the Benjamini-Hochberg procedure \citep{benjamini1995controlling}. We perform these tests on Cohen’s kappa and not on the approximation errors, as the tests require a minimum number of samples.

One-way ANOVAs, separately for each condition, confirm that there are significant differences (all $p<0.01$) in the agreement between participant and metrics (H1).
Paired student’s t-tests in each condition confirm that DreamSim has a significantly higher agreement (all $p<0.01$) than Tile Frequencies (H2). Independent student’s t-tests, followed by p-value correction, confirm that the best metrics from each group, DreamSim (CV), Hamming Distance (General), and Tile Frequencies (PCG), have higher agreement (all $p<0.01$) for pattern-based representations than images (H3). However, this does not hold for all metrics.

\section{Study 2: Interpretation of Similarity Dimensions}
\label{sec:study2}

Our first study has shown that the approximation of human similarity assessments through custom-tailored metrics leaves space for improvements. In our second study, we identified the dimensions underlying the human similarity assessment to better understand this phenomenon and inform the future development of fast and compact similarity metrics tailored to this domain. We thus adopt a similar methodology as in other work on the human perception of similarity (\cref{sec:related_work}) and re-use the perceptual spaces from the first study identified through t-STE on the triplet judgements (\cref{sec:analysis:embedding}), to ask participants in focus groups to interpret their dimensions. To prevent participant fatigue, we employed a mixed design where each condition was assigned to one focus group, tasked to provide interpretations for all four dimensions of the associated perceptual space. We obtained approval from the  \rev{Queen Mary Ethics of Research Committee}.\footnote{\rev{Reference number:
QMERC20.565.DSEECS23.055}}

\subsection{Materials}
We prepared a guide for all participants with a tutorial to demonstrate the exercise (Supplementary Material). It shows a horizontal axis with several circles arranged by increasing size from left to right. The suggested label for this example is “pattern size” or “from small to big”. For each of the four focus groups, we prepare an A2 printout composing all four embedding dimensions of the corresponding condition, to be handed to each participant within (cf.~\cref{appendix:study2:labels}). We leave space under each axis for people to note their ideas. The dimensions are not provided on screen to improve readability and avoid distractions. We used the same demographics and experience questionnaire (Supplementary Material) as in the first study (\cref{sec:study1:materials}) but as a printout.

\subsection{Participants}
Our focus groups were composed of a total of eight 
participants (two per experimental condition) with backgrounds covering HCI and psychology, game AI research, as well as game design and development. These participants were recruited from the \nowordbreak{IGGI} PhD programme, a doctoral training centre spanning multiple universities and focusing on video game research with a strong industry orientation. The study was open for everyone over 18 with normal or corrected to normal vision, which was not assessed. Participants were incentivised with a £15 gift voucher.

Out of the eight participants, seven reported their gender as male, and one as female. The median reported age is 28. Participants in our second study have overall higher experience with the relevant tile-based video games than our general demographic in the first (\cref{tab:gamesexp}).

\subsection{Procedure}
The focus groups were conducted as part of a workshop run at the first author's institution and lasted about 45 minutes each. We ran a total of four individual focus group sessions. All sessions followed the same procedure but focused on interpreting dimensions from different conditions. 

At the beginning of each session, participants were informed about the goals of the study through the participant information sheet (Supplementary Material). They were particularly reminded that multiple interpretations for each dimension are possible, that there are no right or wrong answers, and that their subjective opinion counts. After giving informed consent, they were familiarised with the task through the tutorial sheet and offered help with any remaining questions. They were then handed the sheets with the dimensions to label, one for each participant.

Each session was split into four parts, corresponding to the dimensions on the paper provided to the participants. The experimenter initiated each part by asking the participants to write down their interpretations of the respective dimension silently by themselves. After 5 minutes, they were asked to discuss their proposals with the other members to identify the best interpretation, which they were instructed to write down and highlight. After at most five minutes, the next part was initiated. We decided to interleave the silent individual interpretation task to prevent forgetting about the interpretations and to inspire and inform their upcoming interpretations.

In the debriefing, participants were finally thanked and asked to fill in the demographics and expertise questionnaires. They were then invited to ask any questions, and finally received their incentive, which concluded the session.

\subsection{Results}

\begin{table*}[t]
\caption{Consensus labels for dimensions of the perceptual embeddings (rows) as proposed by individual focus groups per condition (columns) in study 2 (\cref{sec:study2}).}
\label{tab:study2:labels}
\begin{tabularx}{\textwidth}{l|l|l|l|l}
\toprule
Dim. & ccs-img & ccs-pat & loz-img & loz-pat \\
\midrule
1 & \makecell[l]{Shape irregularity (from square blocks\\to non-contiguous shapes)} & \makecell[l]{Pattern complexity (from intricate\\to simple patterns)} & \makecell[l]{Symmetry (from high\\to~low)} & \makecell[l]{Complexity (from\\low to~high)} \\
2 & Level difficulty (from low to~high) & Tile colours (from bright to~dark) & Interesting patterns & — \\
3 & \makecell[l]{Squareness (from vertical/\\horizontal to diagonal shapes)} & \makecell[l]{Pattern symmetry (from vertical\\symmetric to asymmetric)} & \makecell[l]{Colourfulness (from\\low to~high)} & Level theme \\
4 & \makecell[l]{Brightness of tile colours\\(from dark to light colours)} & Tile colours (from blue to~orange) & — & — \\
\bottomrule
\end{tabularx}
\end{table*}

Our findings reflect a diverse range of perspectives, echoing our participants’ varied backgrounds. They noted the difficulty of the labelling task and agreed that discussions within the groups benefited their individual insights. While some groups found it easy to determine consensus labels, not all groups succeeded. 
We compare the consensus labels between conditions in \cref{tab:study2:labels}, and list them, together with the individually assigned labels, under the respective dimension in \cref{appendix:study2:labels}. We summarise the most important findings here.

\paragraph{ccs-img}
The participants believe shape to be of high importance. Their labels include ‘squareness’ (dim. 3) and ‘shape irregularity’ (dim. 1). One participant further mentions ‘roundness’ (dim. 2). The group assigns ‘brightness’ of tile colours as another label (dim. 4).

\paragraph{ccs-pat}
‘Tile colours’, and its range from bright to dark as well as from blue to orange, was assigned as a label to two dimensions (2 and 4). The group further agreed on ‘pattern complexity’ (dim. 1) and ‘pattern symmetry’ (dim. 3).

\paragraph{loz-img}
The group highlighted ‘symmetry’ and ‘colourfulness’ as possible labels (dim. 1 and 3, respectively) and agreed on ‘interesting patterns’ (dim. 2). One participant further mentioned the ‘complexity’ of patterns in relation to two dimensions (2 and 4).

\paragraph{loz-pat}
One participant misinterpreted the tile colours to indicate functionality (blue for water, yellow for desert) and thus focused on game design aspects, describing ‘themes’ of different levels (dim. 3) and the difficulty of solving them. However, the participant also commented on the repetition of tiles (dim. 4), alluding to the distribution of tile types. The group only gave one relevant consensus label: ‘pattern complexity’ (dim. 1).

\section{Discussion}
\label{sec:discussion}

We discuss the findings from our first study (\rev{\cref{sec:study1}}) to contribute to our first research question (\emph{Which existing metrics approximate the human similarity perception of grid-based video game levels best?}). For this, we primarily focus on a metric’s approximation capabilities as quantified by mean squared error in our comparison of similarity matrices, since computing pairwise similarities comes closest to the application scenarios in game development and research. We support these findings with the result from the inter-rater agreement analysis, which allows for a more direct, albeit limited, comparison of the metrics to the participant judgements.

The results suggest that CV-based similarity metrics (CLIP, DreamSim) provide the overall best approximation to the collected participant judgements, outperforming the PCG expert metrics and general-purpose metrics. In particular, results for the artificial neural network-based image embedding DreamSim exhibit the overall lowest approximation error and highest agreement when compared to our participant judgements. While this may be unsurprising, given that the image embedding was specifically fine-tuned to align with human perception of synthetic natural images, our results confirm that this equally benefits similarity estimation of video game levels. Yet, accuracy is not everything. A downside of artificial neural network-based image embeddings however is their size, complexity and dependence on specialised hardware for fast inference. For example, DreamSim requires a CUDA-compatible GPU with 1.6~GB memory available to load the model (additional memory required to compute image embeddings). This can be problematic, considering the limited resources available when relying on such metrics in games at runtime, in particular on mobile devices. Furthermore, sub-symbolic approaches (artificial neural networks) are limited in their transparency, as it is more difficult to explain why a particular pair of levels is attributed to high similarity. In contrast, symbolic approaches (the PCG expert metrics) with their transparent design choices can more easily be broken down into specific rules.

Between the three expert metrics from the PCG literature (with a total of eight configurations), we can observe big differences in performance. The Symmetry metrics in any of its configurations only seem to capture a single aspect relevant to our sample of participants (cf. \cref{tab:study2:labels}), yielding high approximation errors and overall low, often even negative agreement. With the closely related Tile Pattern and Tile Frequencies (identical to the tile pattern size $1\times1$) metrics, we observe a correlation in the results: the larger the patterns, the higher the metric’s approximation error (\cref{fig:study1:sim-mse}). This correlation has a simple explanation: the larger the patterns, the fewer patterns there are in a level to compare. That is to say, a lower granularity of patterns (in the extreme case $1\times1$, i.e. Tile Frequencies) allows for a more nuanced comparison between levels. If there is little data to compare (e.g. only a few large $4\times4$ patterns) it will be difficult to determine whether two levels are slightly more similar than another pair. This can lead to high errors in our similarity matrix analysis. Furthermore, our collection of stimuli is a particularly small dataset, which likely does not provide much overlap in patterns across levels. This explanation is supported by the results on Legend of Zelda levels (\emph{loz-img}, \emph{loz-pat}), which share more patterns due to the common layout of rooms. Consequently, Tile Patterns of size $2\times2$ and $3\times3$ perform much better on levels from this title than on Candy Crush Saga levels. Tile Frequencies being the third-best approximating metric is surprising, given that this metric only compares the number of different tiles in a level but entirely disregards their positions. Nonetheless, our results suggest that Tile Frequencies is a reliable PCG expert metric across all experimental conditions.

The effect of different level structures on metric performance can be observed in the results for Hamming Distance, the best out of two general-purpose metrics. Hamming Distance performs much better on Legend of Zelda than on Candy Crush Saga levels. As mentioned above, the common structure of Legend of Zelda levels puts a focus on the differences in the room interiors between levels. All rooms are the same size and are surrounded by walls and doors. It is thus more important whether rooms are filled with obstacles, enemies or staircases. For participants in our first study, these details may have also been the most similarity-relevant criteria. In contrast, Candy Crush Saga levels can have very different shapes and compositions, making it a more difficult task for a tile position-sensitive metric. Given a more homogeneous collection of Candy Crush Saga levels, Hamming Distance might have performed better on this title. More work is required to test this hypothesis. Hamming Distance has a competitive performance when levels share a common structure and differences between them consist in smaller but important details.

One may argue that in our first study participants with experience of the relevant video game titles (Candy Crush Saga, Legend of Zelda) or similar ones from the same genre might have a better idea of the expressive range amongst levels, therefore making different similarity judgements. Even more so, the perception of expressive ranges between participants, even with similar experiences, might differ. Yet, the design of the triplet judgement task as two-alternative forced choice aims to prevent exactly these variances. Participants are only asked to make a simple binary choice, rather than a more nuanced judgement of similarity.

Our second study (\cref{sec:study2}) allows us to probe this assumption, and highlights two principal similarity-relevant criteria in this specific scenario as an answer to our second research question (\emph{What are the dimensions that govern players’ similarity perception?}). First, the design of patterns in terms of shape (‘irregularity’, ‘squareness’), symmetry and tile composition (‘complexity’). Second, the choice of sprites (‘tile colours’, ‘colourfulness’ and ‘brightness’), which might explain the performance advantage of image embedding metrics, DreamSim and CLIP, over the tile representation-based metric. While symmetry along various axes is already covered by specialised metrics compared in our study, most other criteria are not explicitly accounted for. In particular, visual qualities are not reflected in tile representations. Moreover, participants also thought about gameplay-specific criteria, like level ‘difficulty’ and game narrative (‘themes’), which are not yet covered by any metric. 
All in all, in the context of video games, expert metrics find their purpose in providing robust performance in a dynamic, potentially low-resource environment. These findings can contribute to the future development of custom metrics that meet these requirements and are more closely aligned with human perception.

\subsection{Study Limitations}
\label{sec:limitations}

The present work focuses on visual similarity estimation in two tile-based video games. We note two limitations on \emph{generality}. First, we have not taken into account other game genres beyond tile-based games. Moreover, constrained by the triplet comparison data collection methodology, we could only include a limited amount of stimuli. We tried to mitigate this constraint by systematically selecting stimuli for diversity and through our mixed design. While we selected our two game titles to capture diversity and popularity in the space of tile-based games, there exists much more variation in video game titles that could not be accounted for. Second, the same applies to variation within the levels of each title which, despite our systematic procedure, could not be captured in its entirety. Crucially though, we hold that the dimensions governing similarity between levels here can inform stimulus selection in future studies extending our work. Moreover, we are confident that the choices of games and levels in this work reflect many use cases in the industry.

Beyond limitations to generality, we note that our study only considers similarity judgements of tile-based video games with respect to \emph{visual information}. We agree with related work on player modelling in that functional and dynamic elements of gameplay such as power-ups or tile cascades are also important determinants of player perception, experience and behaviour. Minor differences in the layout of any two levels may have little effect on their visual similarity, yet might make a big difference in terms of gameplay. While many of these elements can be identified visually, we expect players' similarity assessment to be considerably shaped by their active interaction with them. This research thus represents a specialised lens on visual and static game content, contributing to the future development of holistic models of players' similarity judgement.

The setup of experimental conditions and in particular the fact that CV metrics receive different inputs depending on the \emph{Representation} of the condition, limits our study in that we do not cover all possible comparisons for the image-based similarity metrics. We thus do not investigate the discrepancies between the participant judgements across visual representations while keeping the metric representation static. However, we deliver on our plans. As the input to the image-based metrics is varied based on the condition to match what the participants see, we get direct comparisons for how well the image-based metrics approximate the participant judgements for that condition. In this work, we focus on this aspect and leave other comparisons for future work. We acknowledge that the mapping from \emph{img} to \emph{pat} representation does encode some assumptions around the similarity of the different level objects. However, these assumptions do not stem from our own biases but instead rely on the experience of the domain experts for the respective games.

We identify two limitations stemming from the design of the stimuli selection process. First, selecting stimuli that cover the wide range of level designs increases the difficulty of the triplet judgement task. We argue, however, that the data collected from forced-choice judgement tasks is still useful as overall relations between stimuli are captured in the aggregated judgements of a large group of participants. Our results confirm this; despite the difficult triplet combinations, the best metrics compared in our work were able to approximate stimuli relations with very little error. Second, our conclusions about the performance of CLIP are insofar limited as we also \emph{leveraged CLIP in the stimuli selection procedure}. This choice in the selection process allowed us to maximise the diversity of levels, thus benefiting the fair evaluation of all metrics, at the expense of introducing a bias on the performance of a single model. We chose CLIP for the selection procedure as we expected it to be amongst the strongest candidates, thus leaving more space for fine-grained differentiation between the other metrics. 
And despite our use of CLIP in the selection process, our results point to a different CV-based metric as the best-performing metric: DreamSim. We considered multiple other stimuli selection strategies. Here we discuss the advantages and drawbacks of three options which ultimately led us to adopt the approach presented earlier. The first alternative, random sampling of stimuli, is the most unbiased approach, yet unlikely to cover the diverse level design space (e.g. out of 2,792 Candy Crush Saga levels we were only able to select 17). Second, a selection of stimuli informed by a pilot study is also relatively unbiased. However, participants would have to assess an overwhelmingly large amount of stimuli (\emph{ccs}: 2,792 levels; \emph{loz}: 225 levels). A cognitively very demanding task that would require additional recruitment of reliable participants. Third, instead of CLIP, we could use a different embedding (e.g. another CV-based model). While this would not introduce a bias in favour of any of the metrics compared in this work, it would nonetheless introduce a bias towards a different metric for which the relations to the other metrics are not explicitly assessed.

We leveraged inter-rater agreement statistics to facilitate direct comparisons of metric performance with raw data from individual participants. However, we only found \emph{low to moderate agreement between participants}. Since triplet comparisons only require binary decisions, we had no information on participants' confidence in their ratings. We deliberately chose not to leverage disagreement ratios as a proxy for rating confidence due to the low number of samples per stimulus. Future studies could include an additional confidence rating or leverage a different rating task to facilitate a closer comparison between human judgements and the continuous similarity values provided by metrics.

The focus group labelling task in our second study is a naturally noisy process because 1) labels are subjective and 2) the dimensions of the perceptual embeddings are the product of noisy participant judgements. Given the difficulty of the labelling exercise, some groups were not able to provide a consensus label for some dimensions. Yet, rather than labelling exhaustively, our goal was to obtain as much relevant information as possible. The labels identified in our second study nonetheless are a valuable resource to explain the total variance of the similarity judgements. We were only able to obtain data from a relatively \emph{small group of participants} per condition. More participants would have provided higher robustness, as the quality of consensus labels benefits from a variety of perspectives. However, our decision on an on-site study to limit distractions and foster discussion imposed constraints on how many participants could be possibly recruited. Given the complexity of the domain and task, we hold that our findings provide good pointers for future work. Moreover, We published our dataset and interpretation scales to enable other researchers to further validate and extend our findings.

\section{Conclusions \& Future Work}
\label{sec:conclusions}

In this work, we sought to answer two research questions: (1) Which existing metrics approximate the human similarity perception of grid-based video game levels best? And, as a stepping stone toward the development of better metrics, (2) which dimensions govern the similarity perception in this scenario? Of immediate practical relevance, we probe the common belief that the development of good similarity metrics requires a deep understanding of games as an application domain. To this end, we compared 7 metrics in 12 configurations, grouped into custom-made PCG, general-purpose, and computer vision metrics. Surprisingly to some, we find that the DreamSim image embedding exhibits the overall best performance (low overall approximation error and high agreement with human participants), followed by the CLIP embedding model from the same group of CV-based metrics. Since such artificial neural network-based approaches can be too resource-intensive for deployment within a video game, we recommend their use for the offline generation of video game assets.
As an alternative, for in-game use, we find that Tile Frequencies, a simple baseline metric from the PCG literature more suitable for low-resource environments, shows the next-best performance. Furthermore, Hamming Distance is competitive with Tile Frequencies when levels share a common structure and differences between them consist of smaller but important details, e.g. our collection of Legend of Zelda levels.

However, our findings also show that there is room for improvement. Opportunities for advancements of similarity metrics are revealed through our second study, in which we asked focus groups \rev{with relevant experience} to interpret the dimensions underlying the similarity judgement as captured by our data. Participants particularly highlighted the importance of pattern design in terms of shape, symmetry and tile composition, as well as the choice of tile sprites as similarity-relevant criteria of human perception in this specific domain. Our findings contribute to a better understanding of similarity estimation in people and its alignment with existing metrics for tile-based video game levels, and through this inform similarity estimation via computational metrics. 

Our findings can inform metric selection in game development and as an element of research studies on games more generally. Moreover, they highlight potential avenues for improvement of existing and the development of future metrics. We particularly advocate supporting further research on this topic through various uses of machine learning. To select a small subset of stimuli from a large dataset that covers the variation in the dataset, an auto-encoding artificial neural network can be trained on the full dataset. A subset of stimuli can then be selected based on their pairwise distances in the model’s latent space. Other stimuli selection strategies may be applied in future work: random sampling, grid-based selection, etc. To further advance data-driven metrics, we can fine-tune an existing image embedding on a curated dataset of annotated video game levels to obtain a specialised embedding space for the video game domain. Moreover, as DreamSim \citep{fu2023dreamsim} (\cref{sec:related_work}) has demonstrated, we can bootstrap an ensemble of metrics to train a prediction model of human judgement on top of the metrics’ respective calculations. Yet, these efforts have to be assessed in comparison to the performance of much simpler general-purpose metrics. In the video game context, in particular for applications on-device, only limited resources might be available which need to be managed carefully. This work can inform which metrics to include in further benchmarks. We note that this work contributes to the bigger effort of developing holistic models of human similarity judgement in games. Our study setup leaves open for future work further investigations of agreement with CV metric. Our study shows that when participants and image-based metrics are given the same level representation (\emph{img} or \emph{pat}), CV metrics perform best overall. But further study is needed to understand their performance in scenarios where participants are shown the same level screenshots (\emph{img}) between conditions, while the input to image-based metrics is changed from \emph{img} to \emph{pat}. The publication of our data and implementation opens these avenues for future work to the whole research community. More work is needed to extend our analysis to other video game titles, as well as alternative mappings from level objects to abstract colour tiles. While we have focused on the perception of visual similarity in static content, we expect players’ similarity judgement to be also shaped by the dynamic gameplay behaviour that levels afford, and the experiences they are expected to provide. Consequently, an important avenue for future work will be to understand how these static and dynamic aspects can be combined. For example, through representations that can more explicitly encode gameplay, as used in the Video Game Affordances Corpus (VGAC)~\citep{bentley2019videogame}. Finally, while the focus of this paper was on similarity, we advocate research into how well the identified metrics can estimate the human perception of diversity as a natural next step toward supporting a wider range of game AI applications.

Together, our findings can advance a wide range of tasks in research and industry, from developing better player models, more satisfying PCG, believable NPCs, and increasingly plausible automated play-testing approaches. They thus benefit both the game AI and game user research communities and enable new work at the crossroads.

\begin{acks}
We thank our anonymous reviewers for their excellent feedback. 
We thank modl.ai for funding data collection for our first study.
Sebastian Berns was supported by modl.ai and the \nowordbreak{EPSRC} Centre for Doctoral Training in Intelligent Games \& Games Intelligence (\nowordbreak{IGGI}) [EP/S022325/1].
We thank Niki Pennanen for his kind advice on statistics. 
\end{acks}

\newpage

\bibliographystyle{ACM-Reference-Format}
\bibliography{references}

\newpage

\appendix

\section{Study 1: Extended Data Analyses}
\label{appendix:study1}

\rev{There exist some potential limitations in the interpretation of Cohen’s kappa statistic on its own.}
Different scales have been proposed to interpret the magnitude of kappa (e.g. poor, slight, fair, moderate, substantial, and almost perfect; for different intervals of kappa). Yet, choosing any such standard for the evaluation of the strength of agreement is inevitably arbitrary. 
Moreover, a potential scale would have to be adjusted to the maximum value kappa could attain for a given pair of ratings. While $\kappa$ is theoretically upper-bounded by 1, in practice its maximum value is often much lower, as kappa is highly sensitive to differences in allocation and quantities. 
Considering a $2\times2$ contingency table, maximum agreement is only possible if the marginal distributions are balanced. 
We assist the interpretation of kappa by calculating the maximum value of kappa across our pairwise comparisons~\citep{sim2005kappa}, and visualise the difference between individual kappa and their respective maximum values $\kappa_\text{max} - \kappa$ (\emph{unachieved agreement}, lower is better) as raincloud plots in \cref{fig:agreement-diff}. This provides a more realistic scale of comparison across metrics.
We further report two easily interpretable coefficients, appropriate for evaluation of accuracy in prediction tasks, quantity disagreement and allocation disagreement~\citep{pontius2011death}, visualised in \cref{fig:agreement-quant,fig:agreement-alloc}.

\subsection{Results}

\begin{figure*}[h]
  \centering
  \includegraphics[width=\linewidth]{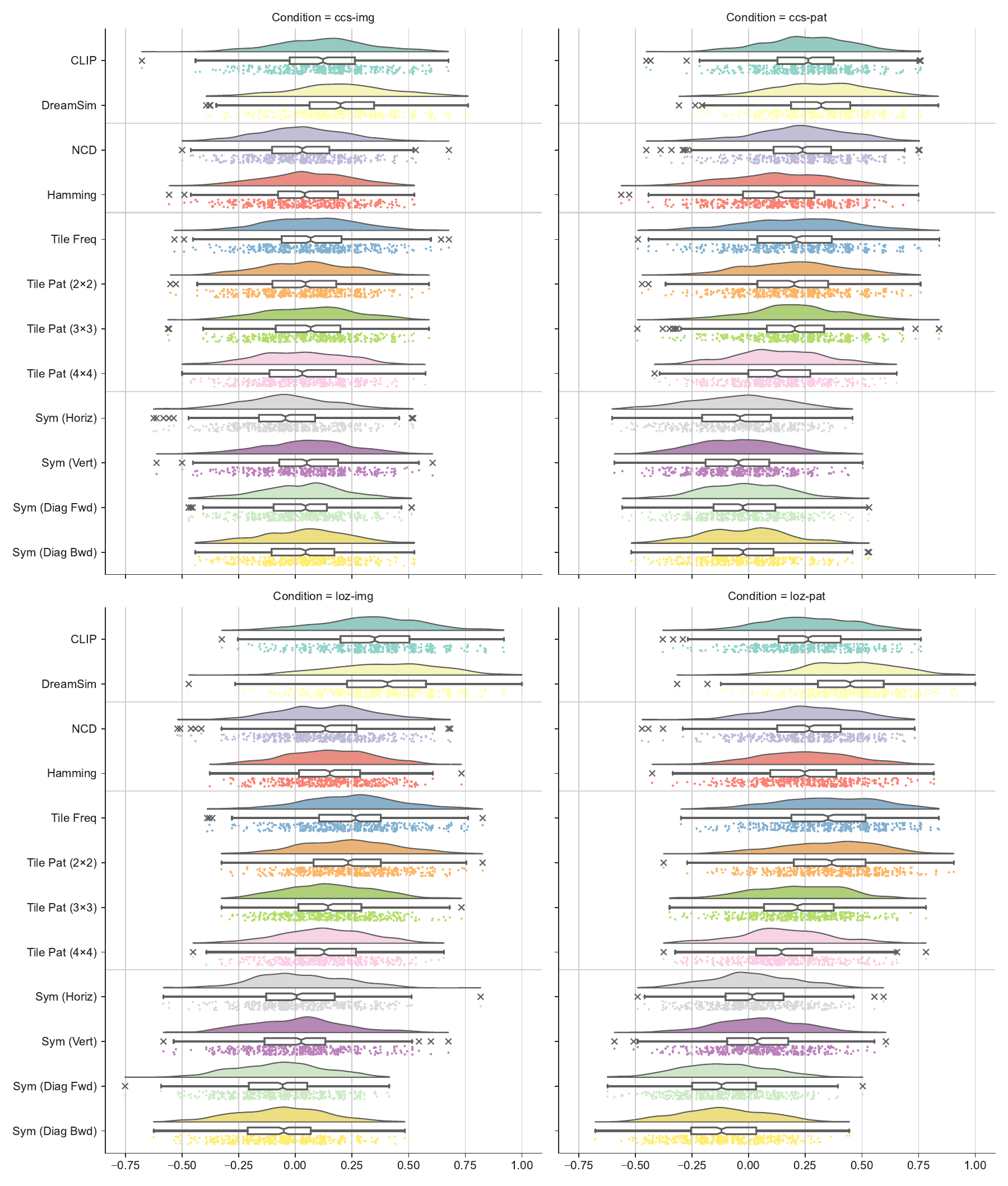}
  \caption{Cohen’s kappa (higher is better): inter-rater agreement between human participants and computational metrics over all experimental conditions (subplots). Each data point indicates Cohen’s kappa comparing the similarity judgements of a single participant against those of a given metric on the same subset of triplets. \rev{Each raincloud plot features individual data points as dots, the estimated kernel density over the data as a curve above the data points, and a box plot with the sample minimum, maximum and median, as well as the first and third quartiles and outliers.}}
  \Description{Four subplots, arranged in two rows and two columns, show results for the different experimental conditions (from top left to bottom right: ccs-img, ccs-pat, loz-img, loz-pat). The subplots share a vertical axis which lists all computational metrics.}
  \label{fig:agreement-kappa}
\end{figure*}

\begin{figure*}[h]
  \centering
  \includegraphics[width=\linewidth]{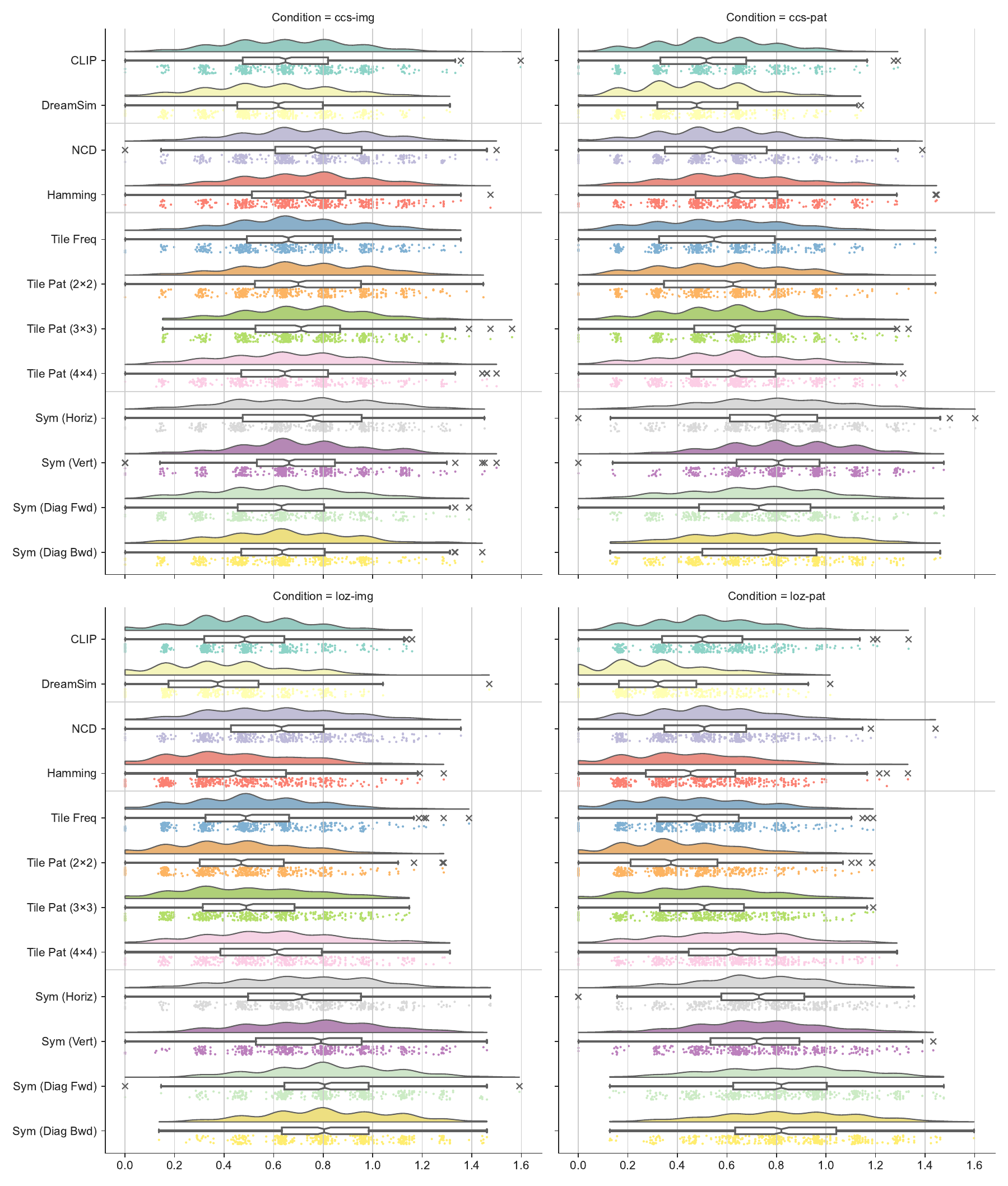}
  \caption{Unachieved agreement (lower is better): difference of the maximum value and Cohen’s kappa of the inter-rater agreement between human participants and computational metrics over all experimental conditions (subplots). Each data point indicates Cohen’s kappa subtracted from $\kappa_\text{max}$, when comparing the similarity judgements of a single participant against those of a given metric on the same subset of triplets. \rev{Each raincloud plot features individual data points as dots, the estimated kernel density over the data as a curve above the data points, and a box plot with the sample minimum, maximum and median, as well as the first and third quartiles and outliers.}}
  \Description{Four subplots, arranged in two rows and two columns, show results for the different experimental conditions (from top left to bottom right: ccs-img, ccs-pat, loz-img, loz-pat). The subplots share a vertical axis which lists all computational metrics.}
  \label{fig:agreement-diff}
\end{figure*}

\begin{figure*}[h]
  \centering
  \includegraphics[width=\linewidth]{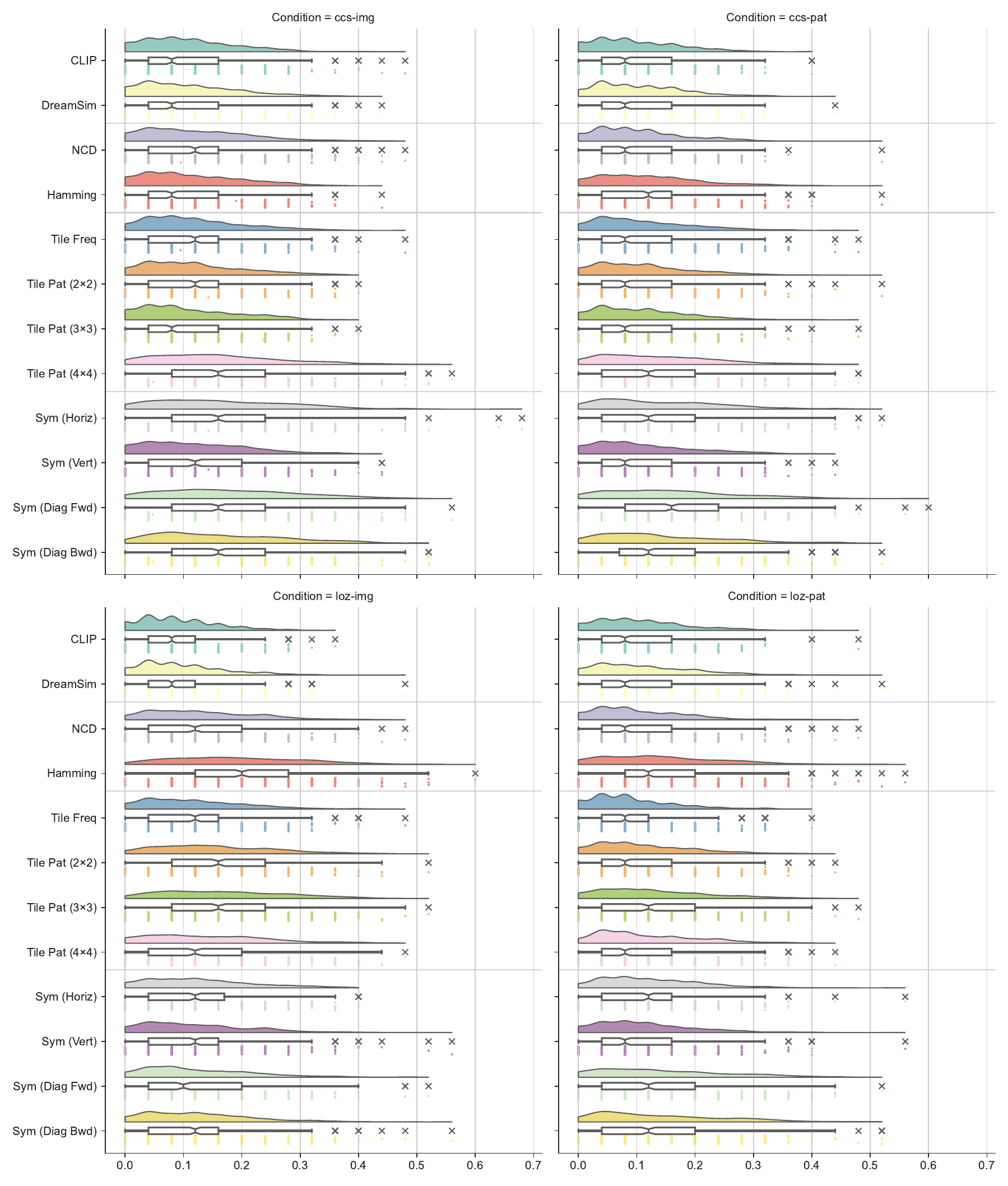}
  \caption{Quantity disagreement (lower is better) between human participants and computational metrics over all experimental conditions (subplots). Each data point indicates disagreement between a single participant and a given metric on the same subset of triplets. \rev{Each raincloud plot features individual data points as dots, the estimated kernel density over the data as a curve above the data points, and a box plot with the sample minimum, maximum and median, as well as the first and third quartiles and outliers.}}
  \Description{Four subplots, arranged in two rows and two columns, show results for the different experimental conditions (from top left to bottom right: ccs-img, ccs-pat, loz-img, loz-pat). The subplots share a vertical axis which lists all computational metrics.}
  \label{fig:agreement-quant}
\end{figure*}

\begin{figure*}[h]
  \centering
  \includegraphics[width=\linewidth]{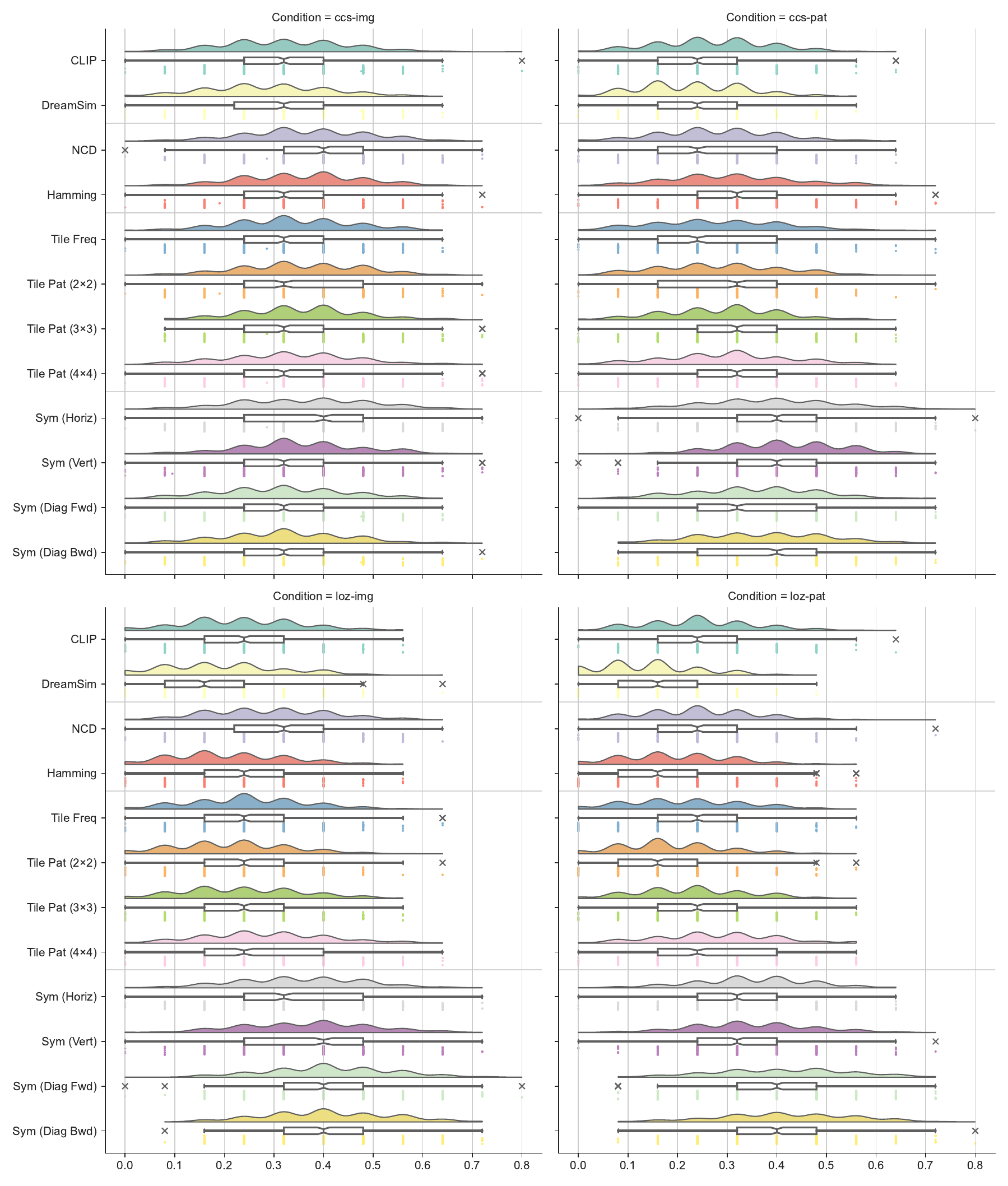}
  \caption{Allocation disagreement (lower is better) between human participants and computational metrics over all experimental conditions (subplots). Each data point indicates disagreement between a single participant and a given metric on the same subset of triplets. \rev{Each raincloud plot features individual data points as dots, the estimated kernel density over the data as a curve above the data points, and a box plot with the sample minimum, maximum and median, as well as the first and third quartiles and outliers.}}
  \Description{Four subplots, arranged in two rows and two columns, show results for the different experimental conditions (from top left to bottom right: ccs-img, ccs-pat, loz-img, loz-pat). The subplots share a vertical axis which lists all computational metrics.}
  \label{fig:agreement-alloc}
\end{figure*}

\rev{We present here the full raincloud plots of Cohen’s kappa from our inter-rater agreement analysis (\cref{fig:agreement-kappa}).}
\rev{To support our main analysis, we report three additional statistics of inter-rater agreement between human participants and computational metrics: (a) unachieved agreement (\cref{fig:agreement-diff}), (b) quantity disagreement (\cref{fig:agreement-quant}), and (c) allocation disagreement (\cref{fig:agreement-alloc}). In all three statistics, lower scores indicate higher agreement. While results are difficult to interpret across all statistics and experimental conditions, there are a few observable patterns, supporting the main analysis.
DreamSim has the overall lowest median scores and interquartile ranges, followed by CLIP and Tile Frequencies. Other metrics occasionally perform better than some of the three but not across all statistics and conditions.}

\section{Perceptual Spaces}
\label{appendix:study2:labels}

This appendix comprises the perceptual spaces obtained via t-STE from triplet comparison data in our first study (\cref{sec:study1}). Each space corresponds to one condition and rests on four dimensions. These were also used as materials in our second, labelling study (\cref{sec:study2}). We include the labels as determined in the study below each dimension. We remind the reader that the study materials were printed in A2 size, thus affording better readability than those presented here.


\begin{figure}[t]
    \centering 
    \includegraphics[width=\linewidth]{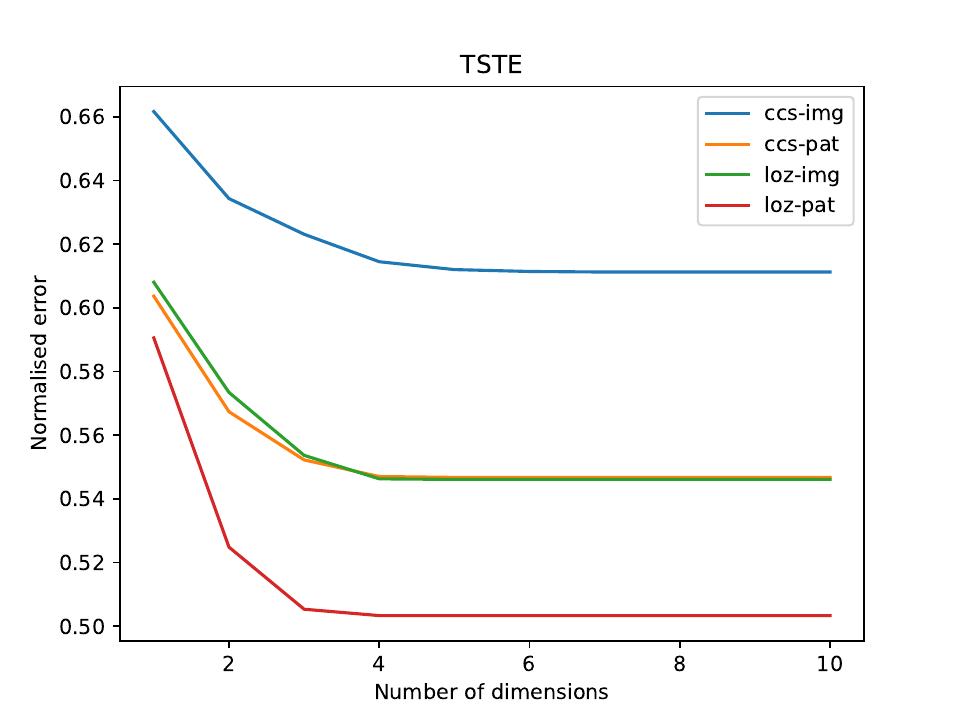} 
    \caption{Elbow plots for t-STE goodness of fit in all conditions (from top left to bottom right: \emph{ccs-img}, \emph{ccs-pat}, \emph{loz-img}, \emph{loz-pat}). We choose 4 as the number of dimensions (horizontal axis) for perceptual embeddings based on the evaluation of overall normalised errors (vertical axis).}
    \Description{Each plot shows a single line that starts with a high error for one dimension, dropping rapidly as the number of dimensions increases. Reaching four dimensions for most conditions (three dimensions for loz-pat), the line flattens out, barely decreasing in error with higher dimensionality.}
    \label{fig:elbows}
\end{figure}


\begin{figure*}[h]
    \subcaptionbox[b]{ccs-img, dimension 1. P1: “From bespoke to generative”. P2: “Irregularity of shapes”. Consensus: “Shape irregularity (from square blocks to non-contiguous shapes)”.}{
        \includegraphics[width=\textwidth]{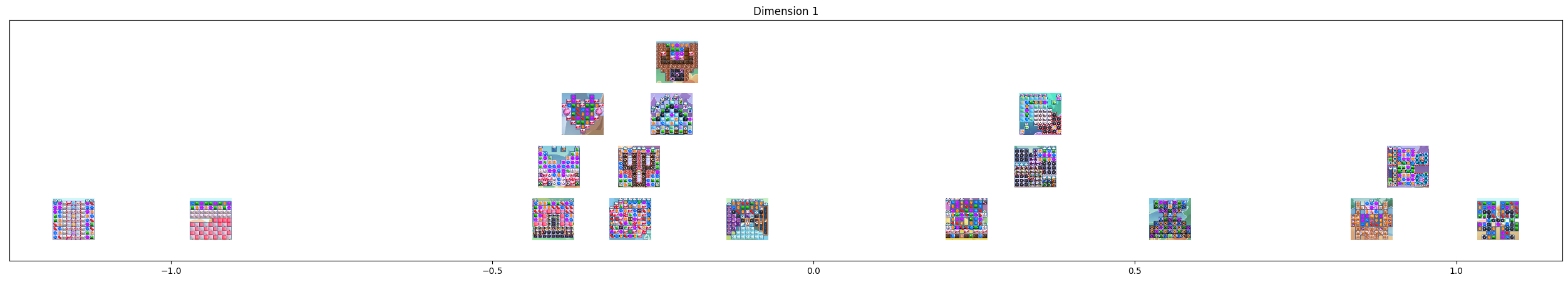}
    }
    \subcaptionbox[b]{ccs-img, dimension 2. P1: “Inverse difficulty (from hard to easy), i.e. more blocks requiring multiple ? (interactions?)”. P2: “Roundness, how much does it look like a circle”. Consensus: “Level difficulty (from low to high)”.}{
        \includegraphics[width=\textwidth]{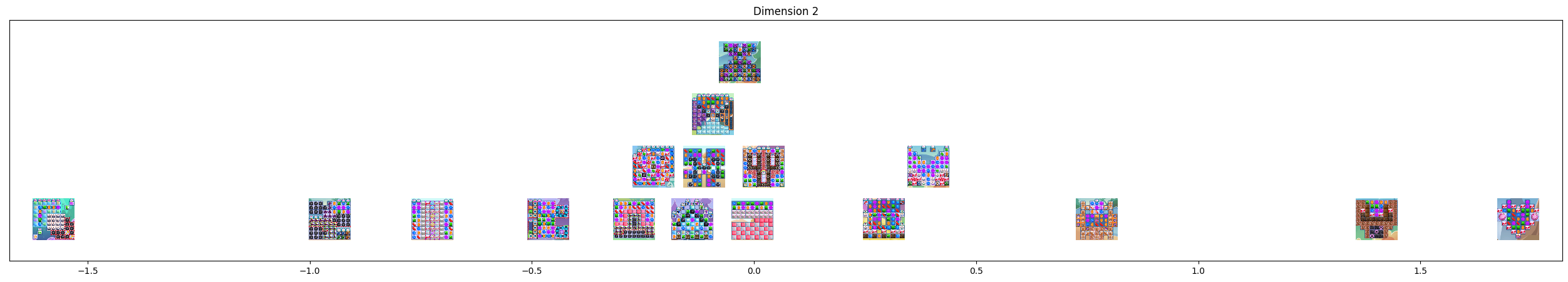}
    }
    \subcaptionbox[b]{ccs-img, dimension 3. P1: “Diagonal angularity (from squareness of level design to diagonalness)”. P2: “Amount of candy/fruit blocks compared to other blocks (just a guess)”. Consensus: “Squareness (from vertical/horizontal to diagonal shapes)”.}{
        \includegraphics[width=\textwidth]{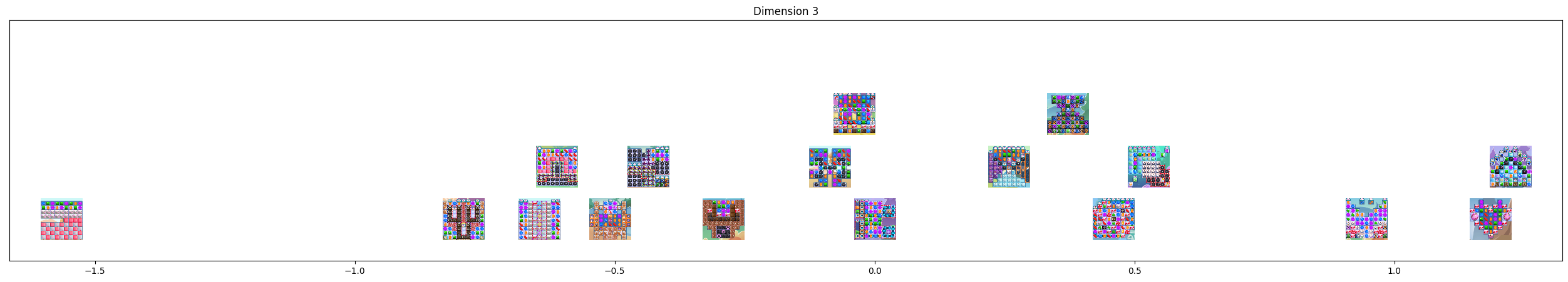}
    }
    \subcaptionbox[b]{ccs-img, dimension 4. P1: “Most to least likely generative (guess)”. P2: “Brightness (from dark to light)”. Consensus: “Brightness of tile colours (from dark to light colours)”.}{
        \includegraphics[width=\textwidth]{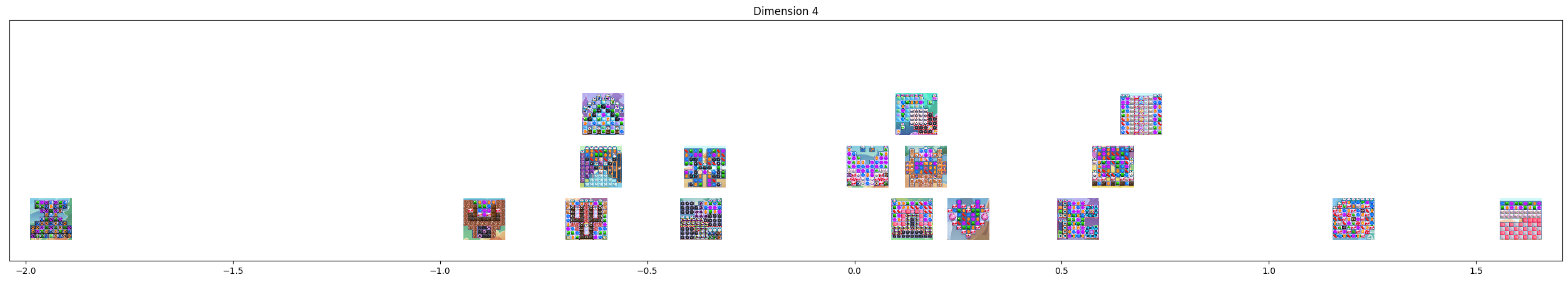}
    }
    \caption{Labelled embedding dimensions for condition \emph{ccs-img}}
    \Description{Screenshots of Candy Crush Saga levels are arranged horizontally by their pairwise similarity. Four subplots visualise the four different dimensions of the perceptual embedding space.}
    \label{fig:study2:labels:ccs-img}
\end{figure*}

\begin{figure*}[h]
    \subcaptionbox[b]{ccs-pat, dimension 1. P3: “Number of straight horizontal lines; pixels are grouped”. P4: “From intricate to simple; Colors tend to loose darker shades from left to right”. Consensus: “Pattern complexity (from intricate to simple patterns)”.}{
        \includegraphics[width=\textwidth]{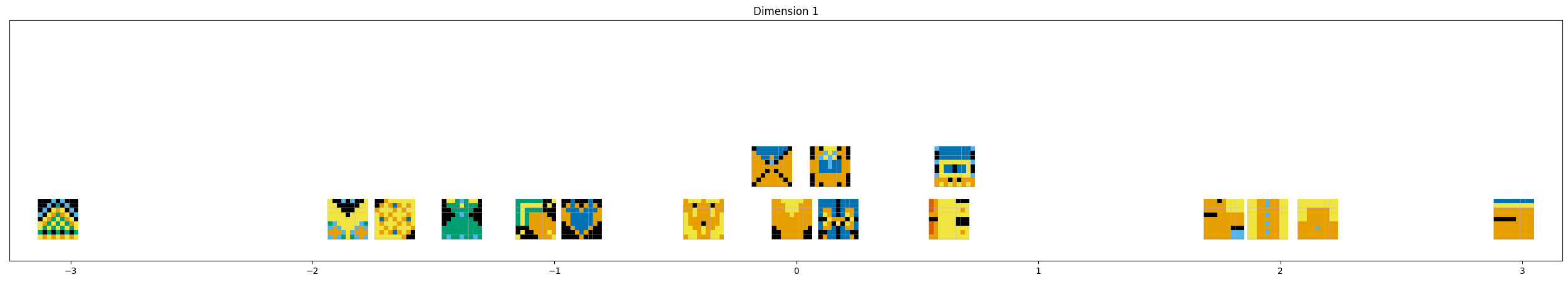}
    }
    \subcaptionbox[b]{ccs-pat, dimension 2. P3: “More green and black, less yellow and blue as x increases”. P4: “Colors tend to go from orange-yellow colorspace to black to green-blue colors (CMY–Black–RGB); Patterns tend too go lateral-symmetric-radial”. Consensus: “Tile colours (from bright to dark)”.}{
        \includegraphics[width=\textwidth]{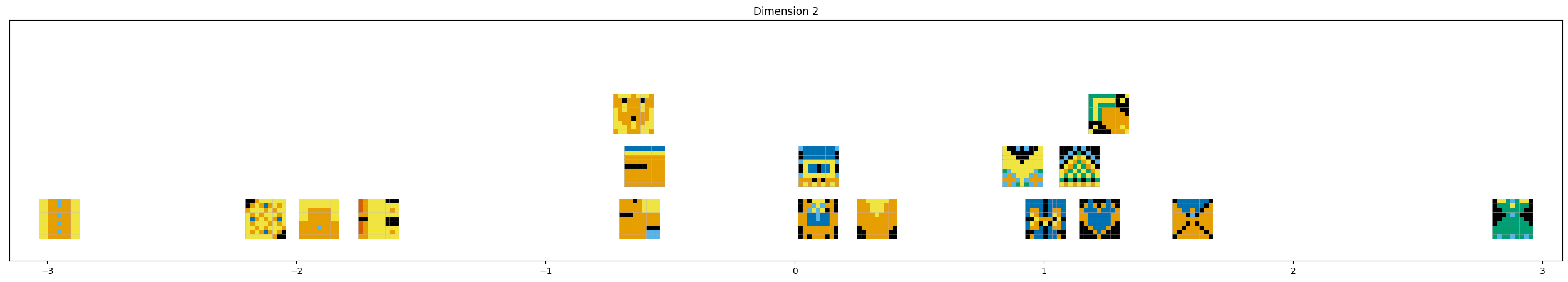}
    }
    \subcaptionbox[b]{ccs-pat, dimension 3. P3: “Blue swaps for green and yellow”. P4: “Pattern from lateral symmetric”. Consensus: “Pattern symmetry (from vertical symmetric to asymmetric)”}{
        \includegraphics[width=\textwidth]{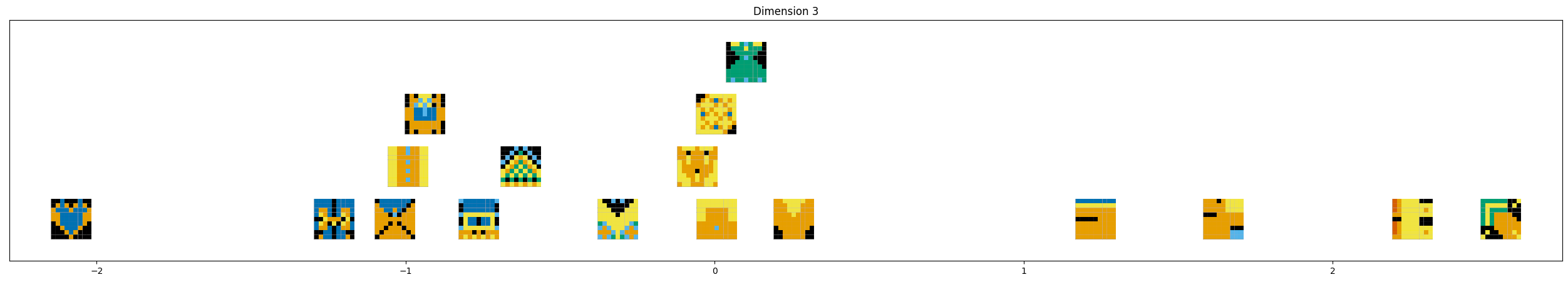}
    }
    \subcaptionbox[b]{ccs-pat, dimension 4. P3: “More orange, less blue as x increases”. P4: “The patterns tend to move up to down going left to right”. Consensus: “Tile colours (from blue to orange)”.}{
        \includegraphics[width=\textwidth]{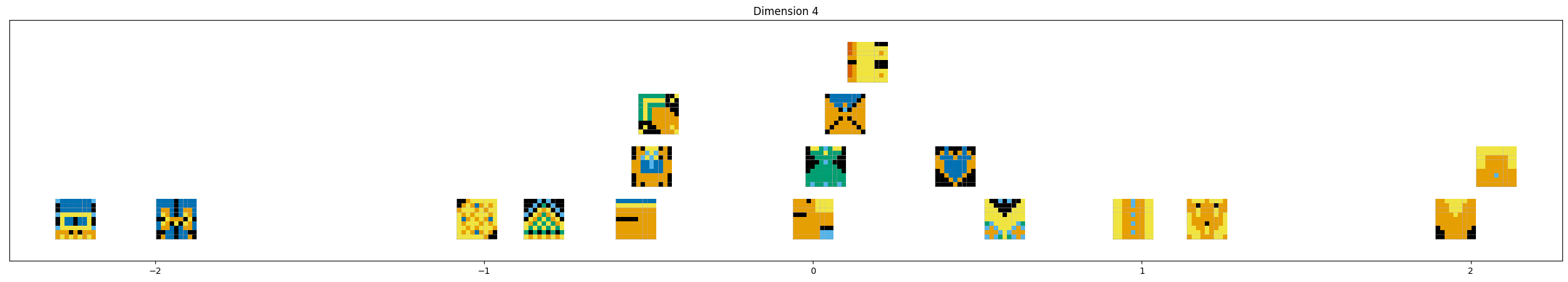}
    }
    \caption{Labelled embedding dimensions for condition \emph{ccs-pat}}
    \Description{Colour pattern representations of Candy Crush Saga levels are arranged horizontally by their pairwise similarity. Four subplots visualise the four different dimensions of the perceptual embedding space.}
    \label{fig:study2:labels:ccs-pat}
\end{figure*}

\begin{figure*}[h]
    \subcaptionbox[b]{loz-img, dimension 1. P5: “Symmetrical arrangement of tiles high – low”. P6: “Asymmetry”. Consensus: “Symmetry (from high to low)”}{
        \includegraphics[width=\textwidth]{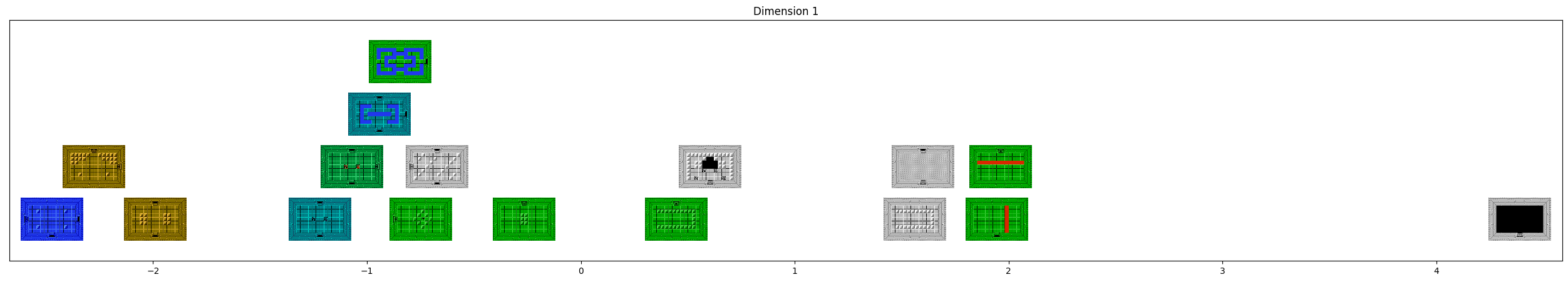}
    }
    \subcaptionbox[b]{loz-img, dimension 2. P5: “Interesting patterns low – high”. P6: “Complexity”. Consensus: “Interesting patterns”.}{
        \includegraphics[width=\textwidth]{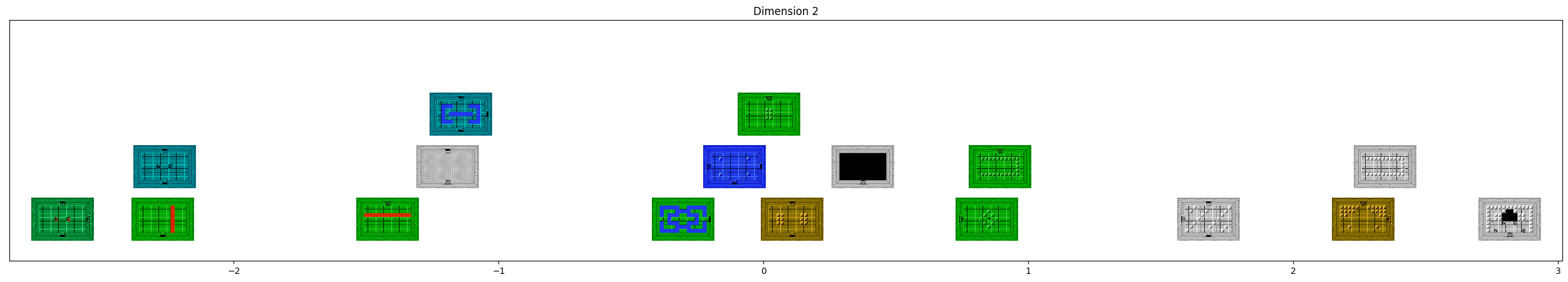}
    }
    \subcaptionbox[b]{loz-img, dimension 3. P5: “Colours variation low – high”. P6: “Incohesion”. Consensus: “Colourfulness (from low to high)”.}{
        \includegraphics[width=\textwidth]{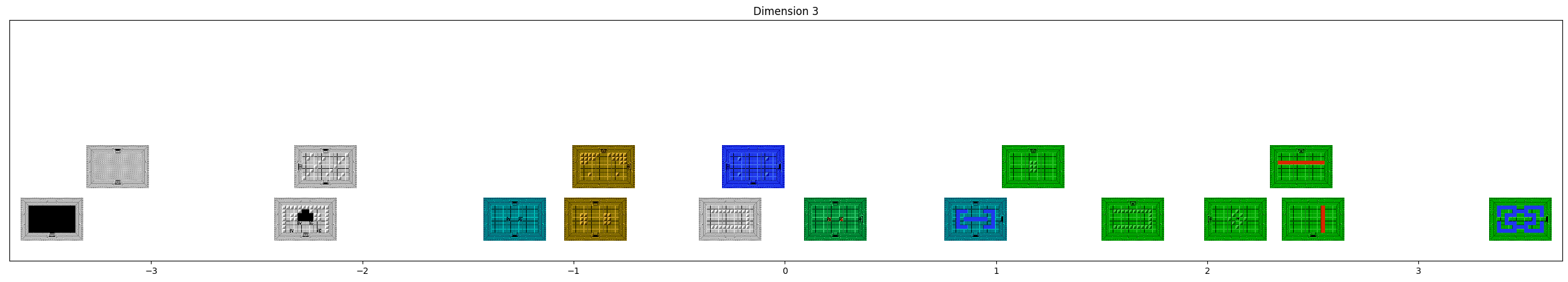}
    }
    \subcaptionbox[b]{loz-img, dimension 4. P5: “Coherence low – high”. P6: “Complexity”. No consensus.}{
        \includegraphics[width=\textwidth]{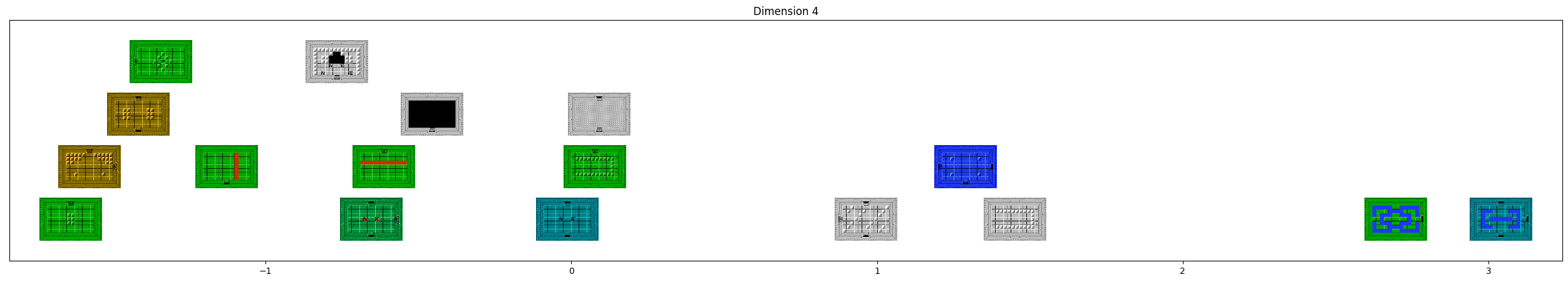}
    }
    \caption{Labelled embedding dimensions for condition \emph{loz-img}}
    \Description{Screenshots of Legend of Zelda levels are arranged horizontally by their pairwise similarity. Four subplots visualise the four different dimensions of the perceptual embedding space.}
    \label{fig:study2:labels:loz-img}
\end{figure*}

\begin{figure*}[h]
    \subcaptionbox[b]{loz-pat, dimension 1.P7: ‘From “game started” to “20 minutes in”’. P8: “Connected components of color (not necessarily of different colors), ignoring outer side rectangles”. Consensus: “Complexity (from low to high)”.}{
        \includegraphics[width=\textwidth]{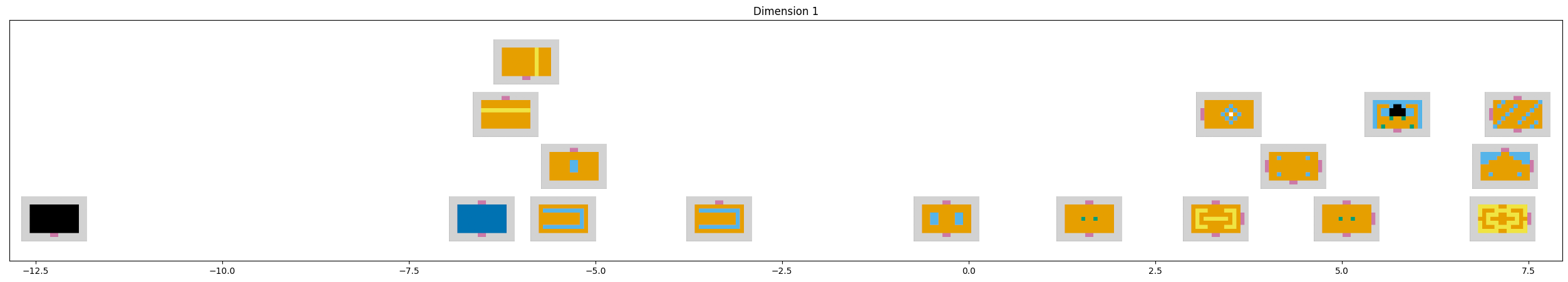}
    }
    \subcaptionbox[b]{loz-pat, dimension 2. P7: ‘I can’t unsee Zelda, so I’m gonna say from “More hidden secrets” –> “less hidden secrets” or “Exploration-focused gameplay” –> “Challenge-focused gameplay”’. P8: “No idea”. No consensus.}{
        \includegraphics[width=\textwidth]{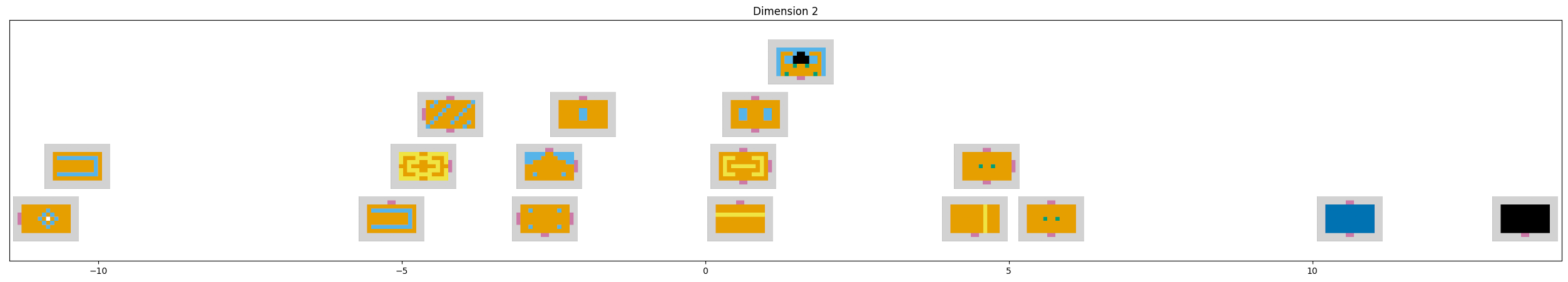}
    }
    \subcaptionbox[b]{loz-pat, dimension 3. P7: ‘“Closed-up areas” –> “Open-ended areas”; Maybe something like “linear progression” –> “Open worlds”; Colour/amount of yellow seems to be a factor too. Maybe “from coast to desert”???; Theme’. P8: ‘Different tile type “theme”; cutscene –> start –> water –> land –> yellow (?)’. Consensus: “Level theme”.}{
        \includegraphics[width=\textwidth]{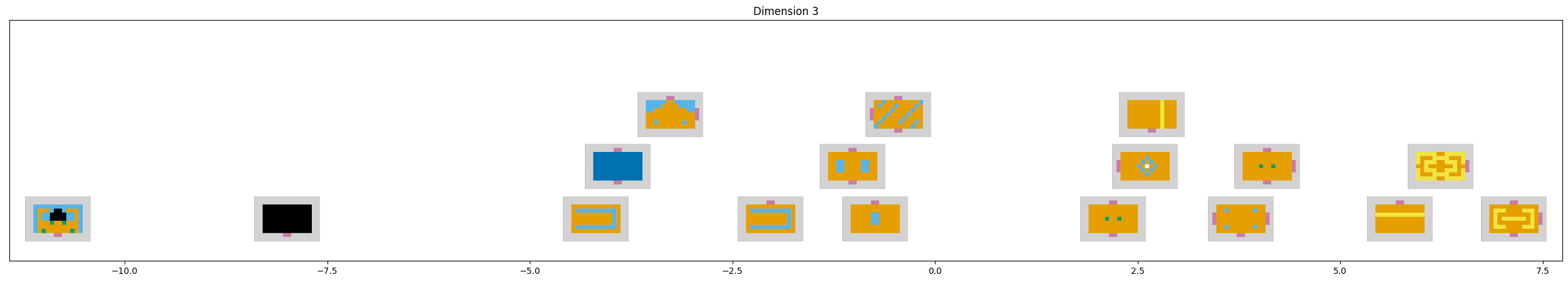}
    }
    \subcaptionbox[b]{loz-pat, dimension 4. P7: ‘More unique to less unique? In the sense of “tile is never repeated in game” –> “tile is often repeated”; Maybe theme again. Yellowish to blueish; Challenging desert section to more relaxed water section. Hard to easy?’. P8: ‘Yellows –> blues’. No consensus.}{
        \includegraphics[width=\textwidth]{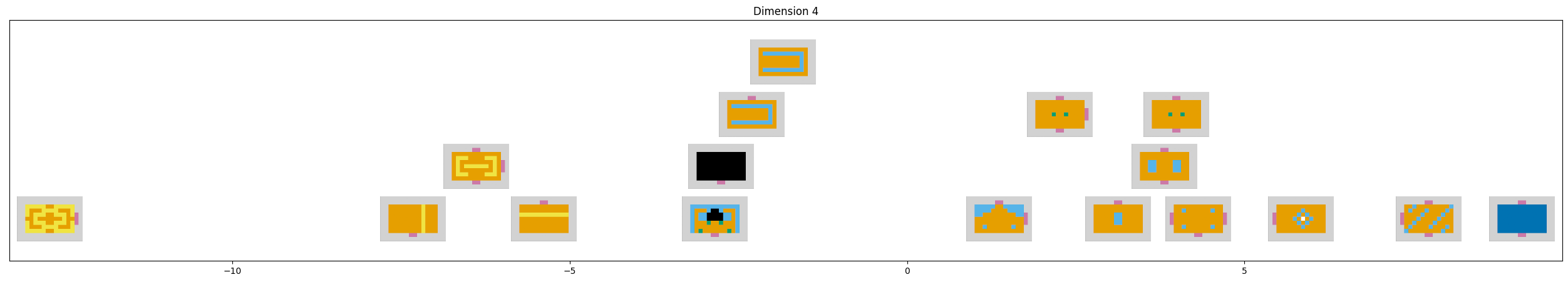}
    }
    \caption{Labelled embedding dimensions for condition \emph{loz-pat}}
    \Description{Colour pattern representations of Legend of Zelda levels are arranged horizontally by their pairwise similarity. Four subplots visualise the four different dimensions of the perceptual embedding space.}
    \label{fig:study2:labels:loz-pat}
\end{figure*}

\end{document}